\documentclass{aa}  

\usepackage{graphicx}
\usepackage{longtable}
\usepackage{lscape}
\usepackage{pdfsync}
\usepackage{bbding}
\usepackage[export]{adjustbox}
\usepackage{natbib} 
\usepackage{txfonts}
\begin{document} 

   \title{On the LINER nuclear obscuration, Compton-thickness and the existence of the dusty torus} 
  \subtitle{Clues from \emph{Spitzer}/IRS spectra} 

   \author{
O. Gonz\'alez-Mart\'in\inst{1,2,3}\thanks{Juan de la Cierva Fellow and Tenure track at CRyA (\email{o.gonzalez@crya.unam.es})}
\and
J. Masegosa\inst{4} 
\and  
I. M\'arquez\inst{4} 
\and 
J.M. Rodr\'iguez-Espinosa\inst{1,2}
\and
J.~A. Acosta-Pulido\inst{1,2} 
\and \\
C. Ramos Almeida\inst{1,2}
\and 
D. Dultzin\inst{5}
\and
L. Hern\'andez-Garc\'ia\inst{4}
\and
D. Ruschel-Dutra\inst{1,2,6}
\and
A. Alonso-Herrero\inst{7,8}
}

   \institute{Instituto de Astrof\'isica de Canarias (IAC), C/V\'ia L\'actea, s/n, E-38205 La Laguna, Spain \and 
Departamento de Astrof\'isica, Universidad de La Laguna (ULL), E-38205 La Laguna, Spain \and 
Centro de Radioastronom\'ia y Astrof\'isica (CRyA-UNAM), 3-72 (Xangari), 8701, Morelia, Mexico \and 
Instituto de Astrof\'isica de Andaluc\'ia, CSIC, Glorieta de la Astronom\'ia s/n 18008, Granada, Spain \and  
Instituto de Astronom\'ia, Universidad Nacional Aut\'onoma de M\'exico, Apartado Postal 70-264, 04510 M\'exico DF, Mexico \and
Departamento de Astronomia, Universidade Federal do Rio Grande do Sul, 9500 Bento Goncalves, Porto Alegre, 91501-970, Brazil \and 
Instituto de F\'isica de Cantabria, CSIC-UC, E-39005 Santander, Spain \and
Department of Physics and Astronomy, University of Texas at San Antonio, San Antonio, TX 78249, USA}

   \date{Received 31st October 2014; accepted 15th January 2015}

 
  \abstract
   {Most of the optically classified low ionisation narrow emission-line regions (LINERs) nuclei host an active galactic nuclei (AGN). However, how they fit into the unified model (UM) of AGN is still an open question. }
   {The aims of this work are to study at mid-infrared (mid-IR) (1) the Compton-thick nature of LINERs (i.e hydrogen column densities of $\rm{N_{H}>1.5\times 10^{24}cm^{-2}}$); and (2) the disappearance of the dusty torus in LINERs predicted from theoretical arguments.}
   {We have compiled all the available low spectral resolution mid-IR spectra of LINERs from the InfraRed Spectrograph (IRS) onboard \emph{Spitzer}. The sample contains 40 LINERs. We have complemented the LINER sample with \emph{Spitzer}/IRS spectra of PG\,QSOs, Type-1 Seyferts (S1s), Type-2 Seyferts (S2s), and Starburst (SBs) nuclei. We have studied the AGN versus the starburst content in our sample using different indicators: the equivalent width (EW) of the polycyclic aromatic hydrocarbon (PAH) at $\rm{6.2 \mu m}$, the strength of the silicate feature at 9.7$\rm{\mu m}$, and the steepness of the mid-IR spectra. We have classified the spectra as SB-dominated and AGN-dominated, according to these diagnostics. We have compared the average mid-IR spectra of the various classes. Moreover, we have studied the correlation between the 12$\rm{\mu m}$ luminosity, $\rm{\nu L_{\nu}(12\mu m)}$, and the 2-10 keV energy band X-ray luminosity, $\rm{L_{X}(2-10~keV)}$.}
   {In 25 out of the 40 LINERs (i.e., 62.5\%) the mid-IR spectra are not SB-dominated, similar to the comparison S2 sample (67.7\%). The average spectra of both SB-dominated LINERs and S2s are very similar to the average spectrum of the SB class. The average spectrum of AGN-dominated LINERs is different from the average spectra of the other optical classes, showing a rather flat spectrum at $\rm{6-28 \mu m}$. We have found that the average spectrum of AGN-dominated LINERs with X-ray luminosities $\rm{L_{X}(2-10~keV)> 10^{41}erg/s}$ is similar to the average mid-IR spectrum of AGN-dominated S2s. However, faint LINERs (i.e. $\rm{L_{X}(2-10~keV)< 10^{41}erg/s}$) show flat spectra different from any of the other optical classes. The correlation between $\rm{\nu L_{\nu}(12\mu m)}$ and $\rm{L_{X}(2-10~keV)}$ for AGN nicely extends toward low luminosities only if SB-dominated LINERs are excluded and the 2-10 keV band X-ray luminosity is corrected in Compton-thick LINER candidates.}
   {We have found that LINERs proposed as Compton-thick candidates at X-ray wavelengths may be confirmed according to the X-ray to mid-IR luminosity relation. We show evidence in favour of the dusty-torus disappearance when their bolometric luminosity is below $\rm{L_{bol}\simeq 10^{42}erg/s}$. We suggest that the dominant emission at mid-IR of faint LINERs might be a combination of an elliptical galaxy host (characterised by the lack of gas), a starburst, a jet, and/or ADAF emission. Alternatively, the mid-IR emission of some of these faint LINERs could be a combination of elliptical galaxy plus carbon-rich planetary nebulae. In order to reconcile the Compton-thick nature of a large fraction of LINERs with the lack of dusty-torus signatures, we suggest that the material producing the Compton-thick X-ray obscuration is free of dust. }


   \maketitle

\section{Introduction}

The emission in active galactic nuclei (AGN) is powered by accretion onto a supermassive black hole (SMBH). AGN are traditionally divided into two main classes based on the presence (Type-1) or not (Type-2) of broad permitted lines (FWHM$\rm{>}$2000 km $\rm{s^{-1}}$) in the optical spectrum. The so-called unification model (UM) proposes that both types of AGN are essentially the same objects viewed at different angles \citep{Antonucci93,Urry95}. An optically thick dusty torus surrounding the central source would be responsible for blocking the region where these broad emission lines are produced (the broad line region, BLR) in Type-2 Seyferts. The torus must not be spherically symmetric, in order to obscure the BLR, while allowing at the same time the emission coming from region producing the permitted narrow lines (known as narrow-line region, NLR) to reach us from the same line of sight (LOS). 

Low-ionisation nuclear emission-line regions (LINERs), first classified by \citet{Heckman80}, are the dominant population of AGN in the local Universe \citep{Ho97}. However, they remain as one of the most captivating subsets of nuclear classes because their main physical mechanism is still unknown. The nature of LINERs was initially sustained in their optical spectrum, which can be reproduced with a variety of different physical processes \citep[e.g. photoionisation from hot stars, non-thermal photoionisation, shocks, post-main sequence stars, or AGN,][]{Dopita95,Heckman80,Ferland83,Veilleux87,Stasinska08,Singh13}. In fact, many authors have suggested that LINERs are an heterogeneous class \citep{Satyapal05,Dudik05,Dudik09}. Indeed the emission mechanism dominating their optical spectrum is still under debate. \citet{Singh13} show that an AGN alone cannot explain the radial profile of the surface brightness H$\rm{\alpha}$ emission line; a contribution of extended emission post-main sequence stars is needed at optical frequencies. The reason why this contribution can be seen in LINERs might be the intrinsic weakness of the AGN, which would outshine these signatures in more powerful AGN. In line with that, \citet{Gonzalez-Martin14} showed that the host-galaxy contributes in a large fraction in most of the LINERs even at X-rays. Thus, the analysis of LINERs could be key to study the interplay between the AGN and the host galaxy.

Nowadays we know that around 75-90\% of LINERs show evidence of AGN using multi-wavelength information \citep{Gonzalez-Martin06,Gonzalez-Martin09A,Dudik09,Younes11,Asmus11,Mason12,Gonzalez-Martin14} \footnote{Note that most of these studies select their sources using archival X-rays observations so they might be biased due to the complex selective effect this introduces in the sample.}. This is confirmed also from X-ray and UV variability studies \citep{Maoz05,Hernandez-Garcia13,Hernandez-Garcia14}. However, what does make LINERs different from other AGN? How do they fit into the UM of AGN? Some results suggest that they might constitute a class of AGN with a different accretion mode \citep[e.g.][]{Younes11,Nemmen14} while some other authors have argued that large obscuration is responsible for their differences \citep[e.g.][]{Dudik09,Gonzalez-Martin09B}. \citet{Gonzalez-Martin09A} found that the hydrogen column density, $\rm{N_{H}}$, in LINERs range from the galactic value up to $\rm{N_{H}\simeq10^{24}~cm^{-2}}$. This is fully consistent with the $\rm{N_{H}}$ values reported for Seyfert galaxies \citep[e.g.][]{Panessa06,Bianchi12,Marinucci12}. However, using the ratio between the luminosity of the [OIII]$\rm{\lambda 5007\AA}$ emission line and the intrinsic hard (2-10 keV) X-ray luminosity, L([OIII])/$\rm{L_{X}(2-10~keV)}$, as a tracer of Compton-thickness (i.e. $\rm{N_{H}>1.5\times10^{24}~cm^{-2}}$), \citet{Gonzalez-Martin09B} found that up to 53\% of the LINERs in their sample are Compton-thick candidates. This percentage is two times higher than that reported for Type-2 Seyferts \citep{Maiolino98,Bassani99,Panessa06,Cappi06}. \citet{Dudik09} studied the emission lines in 67 high-resolution \emph{Spitzer}/IRS spectra of LINERs and found that the central power source in a large percentage of LINERs is highly obscured at optical frequencies, consistent with the X-ray results. 

Obscuring dust hampers the studies of the optical to soft X-ray emission coming from both young hot-stars and the accretion disk. However, emission at mid-infrared (mid-IR) wavelengths does not suffer such a large extinction. Furthermore, the dust that absorbs the shorter wavelength emission reradiates in the mid-IR and correspondingly produces a substantial fraction of the bolometric flux of the object. Dissecting the detailed mid-IR spectra of AGN can reveal the properties of the dust in the nuclear region \citep[e.g.][]{Mendoza-Castrejon15}. Subarcsecond resolution studies claim a tight correlation between rest frame luminosities $\rm{\nu L_{\nu} (12\mu m)}$ and $\rm{L_{X}(2-10~keV)}$ for Type-1 and Type-2 Seyferts \citep{Horst09,Gandhi09,Asmus11,Masegosa13,Gonzalez-Martin13}. Imaging data at mid-IR wavelengths have shown that some Compton-thick LINER candidates might fall off this relation, with $\rm{L_{12\mu m}}$ larger than expected for their $\rm{L_{X}}$ \citep{Mason12,Masegosa13}. The confirmation of such a trend for a large sample might confirm the Compton-thick nature of a large portion of LINERs.  

From the theoretical point of view, \citet{Elitzur06} showed that the torus might disappear when the bolometric luminosity decreases below $\rm{L_{bol}\sim 10^{42} erg/s}$ because the accretion onto the SMBH cannot longer sustain the required cloud outflow rate. Thus, the low bolometric luminosity of LINERs makes them key to probe this theoretical prediction. \citet{Maoz05} showed that the fraction of variable Type-1 and Type-2 LINERs at UV is the same. This favours the lack of a dusty torus obscuring the central AGN in Type-2 LINERs since the central source is the responsible for such variability pattern. The mid-IR emission shows clear signatures of the dusty torus. In fact, clumpy torus models \citep{Nenkova08} have succeeded in explaining the mid-IR emission of Seyfert galaxies \citep[e.g.][]{Ramos-Almeida09,Ramos-Almeida11,Alonso-Herrero11,Honig10}. Thus, the mid-IR spectra of LINERs can give important clues on the existence of the dusty torus for low bolometric luminosities. However, other contributors like jet emission \citep[e.g. NGC\,1052,][]{Fernandez-Ontiveros12} or optically thin dust \citep[e.g. NGC\,3998,][]{Mason13} can also contribute to the mid-IR emission.

The purpose of this paper is to study (1) the Compton-thick nature of LINERs and (2) the plausible disappearance of the torus. We present the mid-IR \emph{Spitzer}/IRS spectra of 40 LINERs. We compare them with mid-IR \emph{Spitzer}/IRS spectra of Seyferts, PG\,QSOs, and Starbursts (see Section \ref{sec:sample}). The data reduction and measurements are described in Section \ref{sec:data}. The relatively low spatial resolution of \emph{Spitzer}/IRS spectra makes these spectra being contaminated from the host galaxy emission, which is particularly relevant for low-luminosity AGN (LLAGN) as LINERs. Section \ref{sec:AGNvsSB} describes a method which is able to select those mid-IR spectra with a negligible starburst contribution. Section \ref{sec:AverageSpectra} studies the average spectrum of LINERs and compare them with that of Seyferts, PG\,QSOs, and Starbursts. Section \ref{sec:XrayMIR} shows the analysis  of the correlation between $\rm{L_{X}(2-10~keV)}$ and $\rm{L_{12\mu m}}$ for LINERs. In Section \ref{sec:discussion} we discuss the implications of the main results. The conclusions of this paper are summarised in Section \ref{sec:conclusions}.

\section{Sample}\label{sec:sample}

Our initial sample of LINERs comes from the catalog of LINERs observed at X-rays published by \citet{Gonzalez-Martin09A}. This guarantees that all the LINERs have $\rm{L_{X}(2-10~keV)}$ measurements, what is crucial for our purposes. However, we must be aware that this sample does not constitute a complete sample. For the present analysis we have used two databases to obtain the \emph{Spitzer} data for an additional sample of LINERs. The first one is the Cornell atlas of \emph{Spitzer}/IRS spectra (CASSIS\footnote{http://cassis.astro.cornell.edu/atlas/}). CASSIS provides low-resolution spectra (R$\rm{\sim}$60-127 over 5.2$\rm{\mu}$m to 38$\rm{\mu}$m) with the IRS instruments in the stare mode \citep{Lebouteiller11}. The second database is the \emph{Spitzer} infrared nearby galaxy survey \cite[SINGS][]{Kennicutt03}. SINGS is a Legacy programme of imaging and spectroscopic data for 75 nearby galaxies. As part of the Legacy programme all the one-dimensional nuclear spectra have been archived in the infrared science archive (IRSA\footnote{http://irsa.ipac.caltech.edu}). This provides uniform 5-30$\rm{\mu m}$ spectra in the spectral mapping mode. Most galaxies in the SINGS sample have also been observed with \emph{Chandra} and their main X-ray properties are described by \citet{Grier11}. 

The CASSIS atlas contains 27 LINERs published by \citet{Gonzalez-Martin09A}. We have also added NGC\,3079, that was not analysed by \citet{Gonzalez-Martin09A} but it was included as a Compton-thick LINER by \citet{Goulding12}. \citet{Grier11} included 20 LINERs \citep[8 included in][]{Gonzalez-Martin09A}. 

The final sample of LINER in this paper contains 40 \emph{Spitzer}/IRS spectra. Among them eight have been optically classified as Type-1.9 LINERs (LINER1) and 32 as Type-2 LINERs (LINER2) by \citet{Ho97}. Three of them are known Compton-thick, 21 are Compton-thin, and 16 were classified as Compton-thick candidates by \citet{Gonzalez-Martin09B}. 

\citet{Dudik09} reported mid-IR spectra of 67 LINERs (13 objects in common with our sample). However, they used the high-resolution modes of \emph{Spitzer}/IRS because their work was focused in the fine structure mid-IR emission lines. Thus, their results are not directly comparable with ours. \citet{Sturm06} reported a mid-IR study of 33 LINERs. They selected their sample on the basis of IR luminosity while our sample is conformed by objects with measured X-ray luminosities. As a consequence, only NGC\,4486 is common with our sample. 

\subsection{Comparison samples}

To study the possible contribution of star-formation or AGN emission, we have selected starbursts, Seyferts and PG\,QSOs to be able to compare their mid-IR spectra to those of LINERs. Note that the sample is not complete in any sense but it allows us to have a representative set of objects for each category to compare LINERs with them:

\begin{itemize}
\item Seyferts. All the Type-1 and Type-2 sources included in \citet{Shi06}, in the Compton-thick sample described by \citet{Goulding12}, and those included in the SINGS sample. In total it contains 42 Seyferts. Among them 31 are Type-2 Seyferts (S2, including 19 Compton-thick and 12 Compton-thin) and 11 Type-1 Seyferts (S1)\footnote{Note that the S1 sample contains objects classified as Type 1, Type 1.2, Type 1.5, Type 1.8, and Type 1.9 Seyferts. The S2 sample includes only purely Type-2 Seyferts.}. 
\item Palomar Green\,QSOs (PG\,QSOs). This sample includes all the PG\,QSOs in the sample defined by \citet{Veilleux09} with \emph{Spitzer}/IRS spectra in CASSIS and redshifts $\rm{z<}$0.2\footnote{The redshift limit has been chosen to be able to obtain rest frame 30$\rm{\mu m}$ luminosities, required for our analysis.}. This PG\,QSO sample includes 26 sources. 
\item Starbursts. This sample is taken from \citet{Ranalli03}, \citet{Brandl06}, and \citet{Grier11}. The starburst sample contains 21 sources. Note that among them NGC\,3367 was classified as a Seyfert by \citet{Veron06} although it was classified as a starburst by \citet{Ho97}. 
\end{itemize}

Note that for all the classes we have only included spectra observed with both the short-low (SL) and long-low (LL) modules to guarantee the full \emph{Spitzer}/IRS coverage (at least $\rm{\sim 5-30 \mu m}$). Moreover, ultra-luminous infrared galaxies (ULIRGs) have been excluded from the analysis because they might have a controversial source of emission at mid-IR \citep[][]{Imanishi07,Alonso-Herrero13}. All together these samples comprise 89 sources (129 nuclei including LINERs). 

\section{Data processing and analysis}\label{sec:data}

CASSIS and SINGS provide flux and wavelength calibrated spectra. However, the observations using data from both the SL and LL spectral modules suffer from mismatches due to telescope pointing inaccuracies or due to different spatial resolution of the IRS orders. This is not corrected in the final products given by CASSIS and SINGS. We therefore scaled each spectra to the immediate prior (in wavelength range) to overcome such effects. Thus, our flux level is scaled to the level of the shortest wavelengths, which is the order with the highest spatial resolution. This guarantees that the flux level is scaled to the best spatial resolution that \emph{Spitzer} can provide. Moreover, the spectra are shifted to rest-frame according to the redshift of the objects.

For each object we have measured the 12$\rm{\mu}$m and 30$\rm{\mu}$m luminosities using the \emph{Spitzer}/IRS spectra. Errors have been estimated assuming 15\% flux-calibration uncertainties, which fully dominate other source of errors \citep[e.g.][]{Gonzalez-Martin13,Ramos-Almeida11}. We have also measured the fluxes and equivalent width (EW) of the polycyclic aromatic hydrocarbon (PAH) features at 6.2 and 11.3 $\rm{\mu m}$. The EW of the PAH features were measured by integrating the emission over the continuum in a wavelength range of 5.9-6.5$\rm{\mu m}$ and 11.0-11.6$\rm{\mu m}$ for the 6.2 and 11.3$\rm{\mu m}$ PAH emission features, respectively. The continuum was estimated through a linear fit to the 5.5-5.9$\rm{\mu m}$ (10.7-11.0$\rm{\mu m}$) and 6.5-6.7$\mu m$ (11.6-11.9$\rm{\mu m}$) for the 6.2$\rm{\mu m}$ (11.3$\rm{\mu m}$) PAH feature \citep[see e.g.][]{Gonzalez-Martin13}. We have also computed the strength of the silicate emission/absorption feature at 9.7$\rm{\mu}$m through the apparent depth at 9.7$\rm{\mu m}$, $\rm{\tau_{9.7\mu m}}$ \citep[e.g.][]{Shi06,Levenson07}:

\begin{equation}
\tau_{9.7\mu m} = ln(F_{cont,9.7\mu m}/F_{9.7\mu m})  
\end{equation}

\begin{figure*}[!t]
\begin{center}
\includegraphics[width=1.\columnwidth]{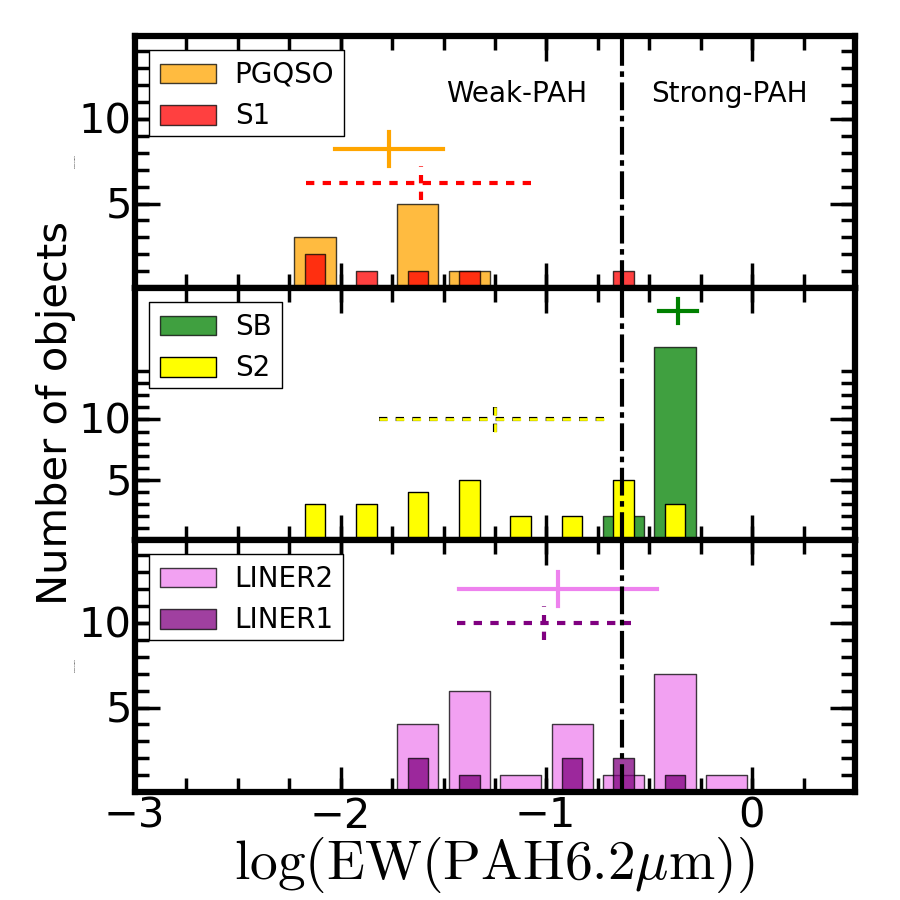}
\includegraphics[width=1.\columnwidth]{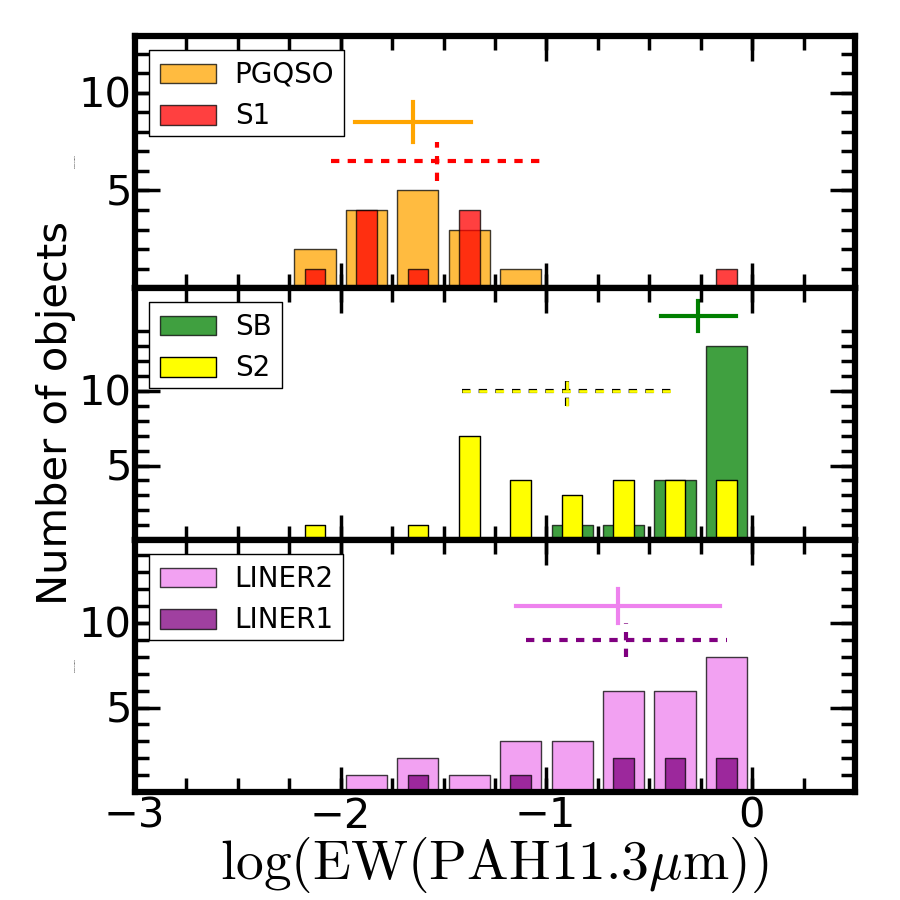}
\caption{Histograms of the EW of the PAH at 6.2 $\rm{\mu m}$ (left) and at 11.3 $\rm{\mu m}$ (right) for PG\,QSO (top-panel, orange-broad bars), S1 (top-panel, red-narrow bars), SB (middle-panel, green-broad bars), S2 (middle-panel, yellow-narrow bars), LINER1 (bottom-panel, purple-narrow bars), and LINER2 (bottom-panel, pink-broad bars). The median values and 25th-75th percentile range for each class of object are shown with large crosses (with the same color-code than the histogram) with continuous lines for PG\,QSO, SB, and LINER2 and with dashed lines for S1, S2, and LINER1. The vertical dot-dashed line shows the limit chosen to divide the sample into strong- and weak-PAHs (see text). }
\label{fig:histoPAH}
\end{center}
\end{figure*}

\noindent where $\rm{F_{9.7\mu m}}$ and $\rm{F_{cont,9.7\mu m}}$ are the fluxes of the spectra around 9.7$\rm{\mu}$m and its expected continuum, respectively. Note that the apparent depth at 9.7$\rm{\mu m}$ $\tau_{9.7\mu m}$ is positive for absorption silicate features and negative for emission features. 

Due to the complexity of \emph{Spitzer} spectra, we have used PAHFIT\footnote{http://tir.astro.utoledo.edu/jdsmith/research/pahfit.php} to obtain $\tau_{9.7\mu m}$ except when emission silicate features are detected (see below). PAHFIT is an IDL tool for decomposing \emph{Spitzer}/IRS spectra of PAH emission sources, with a special emphasis on the careful recovery of ambiguous silicate absorption, and weak, blended dust emission features \citep{Smith07}. PAHFIT is primarily designed for its use with the full 5-35$\rm{\mu}$m \emph{Spitzer}/IRS low-resolution spectra. However, PAHFIT is not able to treat or recover silicate emission features expected to occur in Type-1 AGN, giving $\tau_{9.7\mu m}=0$. In these cases we have computed $\tau_{9.7\mu m}$ by fitting the 9-14$\rm{\mu}$m \emph{Spitzer}/IRS spectra to a Gaussian profile. 
This is a general treatment to estimate $\tau_{9.7\mu m}$ that has been proven to be a good approximation when compared with PAHFIT \citep[see][]{Gonzalez-Martin13}.
 
Together with the \emph{Spitzer}/IRS spectra of the sample we also compiled the 12$\rm{\mu}$m luminosities obtained with ground based telescopes. These measurements have the advantage to better isolate the nuclear source because they come from images with $\rm{\sim 0.3}$ arcsec spatial resolution (i.e. few tenths of parsecs for nearby galaxies). Most of these measurements come from the catalog of sub-arcsecond mid-IR images of AGN reported by \citet{Asmus13}. It includes Subaru/COMICS, VLT/VISIR, Gemini/MICHELLE and Gemini/T-ReCS mid-IR data of 253 objects. 

We have also included four LINERs observed with GTC/CanariCam as proprietary data (program IDs GTC42-12B and GTC35-13A). We have reduced them uniformly with the RedCan package \citep{Gonzalez-Martin13}. We present here their luminosities (marked with asterisks in Col. 5 in Table \ref{tab:sample}) whereas the full imaging analysis will be presented in a forthcoming paper. All together we have ground-based measurements for 61 out of the 129 sources included in this paper.

Table \ref{tab:sample} contains the X-ray luminosities, mid-IR measurements for the \emph{Spitzer}/IRS spectra, and the 12$\rm{\mu}$m luminosities from \emph{Spitzer}/IRS and ground-based telescopes when available.

\section{AGN versus starburst contents}\label{sec:AGNvsSB}

\emph{Spitzer} has been used to study the largest samples of AGN ever analysed at mid-IR \citep[e.g.][]{Shi06,Deo07}. However, a disadvantage of these data is their relatively low spatial resolution. This makes \emph{Spitzer} spectra to be often contaminated by the host galaxy. It is expected to be particularly relevant for LINERs where the AGN is faint. In this case the non-AGN contribution might be very strong at mid-IR wavelengths, dominating the entire emission \citep{Mason12}. 

Over the last decade several diagnostics have been proposed to quantify the contribution of star formation and AGN activity to the infrared luminosity. These diagnostics are based on the mid-IR continuum slope, the EW of the PAH features, the ratio of [NeV] (or [OIV]) over [NeII], and the EW of the PAH at 6.2 or 11.3$\rm{\mu m}$ versus the 9.7$\rm{\mu m}$ optical depth $\rm{\tau_{9.7\mu m}}$ \citep{Genzel98,Lutz98,Dale06,Sturm06,Spoon07,Baum10,Hernan-Caballero11}. In this section we use several diagnostics to separate which \emph{Spitzer}/IRS spectra are strongly contaminated by non-AGN emission. Note, however, that the starburst contribution to the mid-IR spectra does not exclude the presence of an AGN. This is an attempt to determine whether the AGN is dominating or not the mid-IR spectrum. We have excluded from the analysis the diagnostics based on fine structure emission lines because they are blended with other emission lines at the spectral resolution of these dataset. For more details in these diagnostics we refer the reader to \citet{Dudik09},  where they studied these emission lines for a large sample of LINERs.

\begin{figure*}[!t]
\begin{center}
\includegraphics[width=2.\columnwidth]{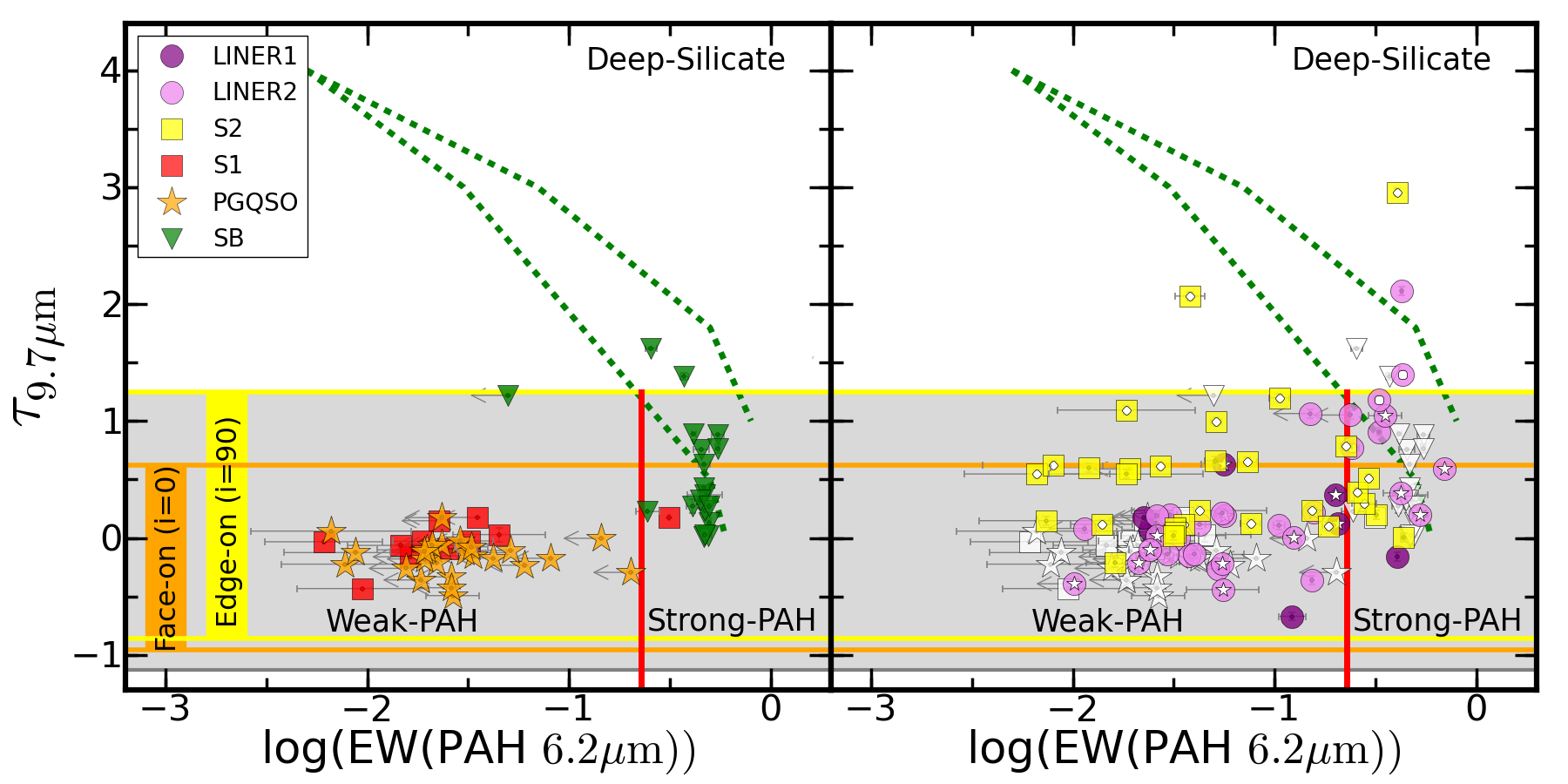}
\caption{The silicate apparent depth at 9.7 $\rm{\mu m}$, $\rm{\tau_{9.7}}$, versus the EW of PAH at 6.2 $\rm{\mu m}$, EW(PAH 6.2 $\rm{\mu m}$), for PG\,QSOs, S1s, and SBs (left) and for LINERs and S2s, (right). LINER1s, LINER2s, S1s, S2s, PG-QSOs, and starbursts are shown with purple and pink circles, red and yellow squares, orange stars, and green up-side down triangles, respectively. White circles and white stars mark known Compton-thick sources and Compton-thick candidates, respectively. Note that the error bars in $\rm{\tau_{9.7}}$ are always within the size of the symbol. In the right panel we also include PG\,QSOs, S1s, and SBs with white symbols for comparison purposes. Green-dotted lines indicate the diagonal branch found by \citet{Spoon07} for ULIRGs and Starbursts. The shadowed (grey) area shows the range of values for $\rm{\tau_{9.7}}$ that could be explained with Clumpy models \citep{Nenkova08}. We also show with orange and yellow lines the expected range of values for $\rm{\tau_{9.7}}$ in face-on AGN (assuming $\rm{i=0^\circ}$) and in edge-on AGN (assuming $\rm{i=90^\circ}$) using the models described by \citet{Nenkova08} (see text). The red-short vertical line shows EW(PAH 6.2$\rm{\mu m)=0.228\mu m}$ which divides into weak and strong PAHs.}
\label{fig:PAH62vsTau}
\end{center}
\end{figure*}

\subsection{PAH features}

The star formation activity correlates with the PAH strength, where starburst-dominated galaxies are then expected to show strong PAH features \citep[e.g.][]{Peeters04}. This well stablished correlation has led to the use of PAH strength as a tracer of star formation \citep[e.g.][]{Esquej14}. PAHs might be destroyed due to the presence of an AGN \citep[][]{Genzel98,Wu09}. This is particularly relevant for the PAH feature at 6.2$\rm{\mu m}$ that is produced by grains with smaller sizes and, therefore, their destruction near the AGN is more efficient \citep[][]{Diamond-Stanic12}. However, the 11.3$\rm{\mu m}$ might not be suppressed by the AGN (at distances as close as 10 pc) but diluted when the AGN continuum becomes dominant \citep[][]{Alonso-Herrero14,Ramos-Almeida14}. Note that our aim is to select those spectra where the host galaxy contribution due to star-formation is not dominating the mid-IR spectrum. The PAH strength is a good tracer of star-formation occurring at far distances from the AGN, where this destruction/dilution of the PAH features is negligible. Supporting this, the PAH at 11.3$\rm{\mu m}$ was negligible in 18 out of the 20 AGN reported by \citet[][]{Gonzalez-Martin13} with high-spatial resolution spectra while their \emph{Spitzer} spectra showed strong PAH features. Thus, these diagnostics are still useful in our analysis, irrespective of the dilution or suppression of the PAH features near the AGN. 

Fig. \ref{fig:histoPAH} shows the distributions of the EW(PAH) per class for the PAH features detected at 6.2$\rm{\mu m}$ (left) and 11.3$\rm{\mu m}$ (right). Note that upper limits on the non detected lines are not included in these histograms. The median EW(PAH 6.2$\rm{\mu m}$) for the SB class is significantly higher than that for the S1 and PG\,QSO classes. Only one S1 shows an EW(PAH 6.2$\rm{\mu m}$) consistent with the SB class (NGC\,5033). Moreover, only one SB shows a limit on the EW(PAH 6.2$\rm{\mu m}$) consistent with S1s or PG\,QSOs (NGC\,3184). The S2 class have objects with EW(PAH 6.2$\rm{\mu m}$) overlapping with values found for S1s, PG\,QSOs, and SBs. LINERs (both LINER1 and LINER2), like S2s, also show EW(PAH 6.2$\rm{\mu m}$) spreading a large range of values. The histogram of EW(PAH 11.3$\rm{\mu m}$) is similar to that of EW(PAH 6.2$\rm{\mu m}$). However, there is a larger overlapping between the distributions in EW(PAH 11.3$\rm{\mu m}$) for SBs with S1s and PG\,QSOs than in the histogram of EW(PAH 6.2$\rm{\mu m}$). Moreover, the S2 class overlaps with the SB class for a larger number of objects in EW(PAH 11.3$\rm{\mu m}$) than for the histogram of EW(PAH 6.2$\rm{\mu m}$) (15 and 6 objects, respectively). 

The PAH feature at 11.3$\rm{\mu m}$ might be strongly attenuated by the silicate absorption feature \citep[][]{Brandl06}. \citet{Gonzalez-Martin13} estimated this attenuation to be up to $\rm{\sim}$40\% of the intrinsic PAH feature at 11.3$\rm{\mu m}$ for $\rm{\tau_{9.7\mu m}= 1}$. This percentage can be higher for larger $\rm{\tau_{9.7\mu m}}$. The EW(PAH 6.2$\rm{\mu m}$) might be a better tracer of star-formation when the silicate attenuation is large because it is not embedded in the silicate absorption feature. This is clearly seen in Fig. \ref{fig:histoPAH}, where the overlap of the S2 class (expected to be more attenuated than S1s under the unified model) with the SBs is much higher for the EW(PAH 11.3$\rm{\mu m}$) than for the EW(PAH 6.2$\rm{\mu m}$). We therefore have chosen EW(PAH 6.2$\rm{\mu m}$) as a better tracer of the SB-dominance in our sample.

We have defined a limit on EW(PAH 6.2$\rm{\mu m}$) using the mean value and the standard deviation over this mean value for objects classified as SBs as follows: $\rm{<EW(PAH6.2\mu m)>-3\times \sigma(EW(PAH6.2\mu m))=0.233 \mu m}$ (i.e. $\rm{<log(EW(PAH6.2\mu m))>=-0.633}$). This ensures that 99.7\% of SBs show EW(PAH 6.2$\rm{\mu m}$) above this limit. Note here that the limit would be  $\rm{EW(PAH6.2\mu m)=0.247}$ if the 99th percentile is used. This will give a slightly less restrictive limit\footnote{{Only two objects will be included in the weak-PAH class if the limit is set to $\rm{EW(PAH6.2\mu m)=0.247}$ compared to those obtained using $\rm{EW(PAH6.2\mu m)>0.233}$, namely NGC\,7130 and NGC\,3367.}}. Above (below) this value we classified the objects as strong-PAH (weak-PAH) objects. Six out of the 31 S2s are consistent with the strong-PAH category;
ten out of the 40 LINERs are classified within the strong-PAH class. All of them are LINER2s except NGC\,1097.

\citet{Spoon07} presented a mid-IR diagnostic of the AGN/ULIRG content based on $\rm{\tau_{9.7}}$ versus EW(PAH 6.2$\rm{\mu m}$) \citep[see also][]{Hernan-Caballero11}. The advantage of this diagnostic is that it takes into account the effects of strong obscuration of the nuclear source. They showed that galaxies are systematically distributed along two different branches: (1) a horizontal line with $\rm{\tau_{9.7}<1}$ of continuum AGN-dominated to PAH-dominated spectra and (2) a diagonal line going from deeply obscured (high $\rm{\tau_{9.7}}$ and low EW(PAH 6.2$\rm{\mu m}$) to PAH-dominated spectra (low $\rm{\tau_{9.7}}$ and high EW(PAH 6.2$\rm{\mu m}$). Seyferts and QSOs are found exclusively on the horizontal branch with $\rm{\tau_{9.7}<1}$. The large majority of LIRGs and ULIRGs in \citet{Spoon07} are located in the diagonal line. Starburst are placed at the end of the two branches, with large EW(PAH 6.2$\rm{\mu m}$) and $\rm{\tau_{9.7}<1}$. They argued that these two branches reflect a fundamental difference in the dust geometry in the two set of sources. The horizontal branch could have a clumpy structure while the diagonal might be smooth. 

Fig. \ref{fig:PAH62vsTau} shows $\rm{\tau_{9.7}}$ versus EW(PAH 6.2$\rm{\mu m}$) for SBs, S1s, and PG\,QSOs in the left panel and LINERs and S2s in the right panel. PG\,QSOs and S1s ($\rm{\tau_{9.7}<1}$ and EW(PAH 6.2$\rm{\mu m )<0.228\mu m}$) are clearly distinguished from SBs (EW(PAH 6.2$\rm{\mu m )>0.233\mu m}$). This result is fully consistent with that reported by \citet{Spoon07}. Our diagram shows very few nuclei with deep silicate features and weak PAH features. This was also found by \citet{Spoon07} with only eight over the 160 objects in their sample belonging to this category. We have tested the use of the EW(PAH 11.3$\rm{\mu m}$) instead of EW(PAH 6.2$\rm{\mu m}$) in this diagram, finding a similar result. This was already reported by \citet{Hernan-Caballero11} in a large sample of \emph{Spitzer}/IRS spectra.

There is a maximum $\rm{\tau_{9.7}}$ expected under the predictions of the clumpy torus models for AGN. Larger values of $\rm{\tau_{9.7}}$ can be interpreted as significant contamination from the host galaxy \citep{Alonso-Herrero11,Gonzalez-Martin13}. In order to investigate this issue, we have computed $\rm{\tau_{9.7}}$ using a set of models within the libraries of CLUMPY\footnote{http://www.pa.uky.edu/clumpy/}. These consist on a set of spectral energy distributions (SEDs) using the AGN clumpy torus emission described by \citet{Nenkova08}. The parameter ranges chosen are those reported by \citet{Gonzalez-Martin13}. We have downloaded the SEDs for a width of the toroidal distribution $\rm{\sigma=45^\circ}$, a ratio between the outer and the inner radius of the torus $\rm{r_{out}/r_{int} = 200}$, an exponential slope of the radial distribution of clouds $\rm{q=2}$, an optical extinction of the clouds ranging $\rm{\tau_{V}=5-150}$, and a number of clouds along the equator of the torus $\rm{N_{o}=2-20}$ clouds. The number of clouds along the LOS, $\rm{N}$, depends on the inclination angle, $i$, as $\rm{N=N_{o}e^{-i^{2}/\sigma^{2}}}$. We refer the reader to \citet{Gonzalez-Martin13} for details on the selection of these parameters and to \citet[][]{Nenkova08} for the details on the modelling. 

We have computed $\rm{\tau_{9.7}}$ for these SEDs using the same methodology as for the \emph{Spitzer}/IRS spectra reported here. Fig. \ref{fig:PAH62vsTau} shows as a grey area the minimum and maximum $\rm{\tau_{9.7}}$ found using these models ($\rm{-1.13<\tau_{9.7}<1.25}$). Thus, objects with $\rm{\tau_{9.7}>1.25}$ are not expected under any clumpy torus model. We use $\rm{\tau_{9.7}>1.25}$ to classify an object as deep-silicate. We also show the range of $\rm{\tau_{9.7}}$ expected for face-on AGN assuming $i=0^{\circ}$ ($\rm{-0.96<\tau_{9.7}<0.62}$) and for edge-on AGN assuming $i=90^{\circ}$ ($\rm{-0.86<\tau_{9.7}<1.25}$). S1s are naturally explained within the range of values of $\rm{\tau_{9.7}}$ expected under the clumpy torus models. S2s tend to show larger $\rm{\tau_{9.7}}$ than S1s. Only two S2s are out of the expected range with clumpy torus models.  
Only two SBs 
and two LINERs 
show $\rm{\tau_{9.7}>1.25}$. 
Following this diagram, we have divided our sample into three categories:

\begin{figure}[!t]
\begin{center}
\includegraphics[width=1.\columnwidth]{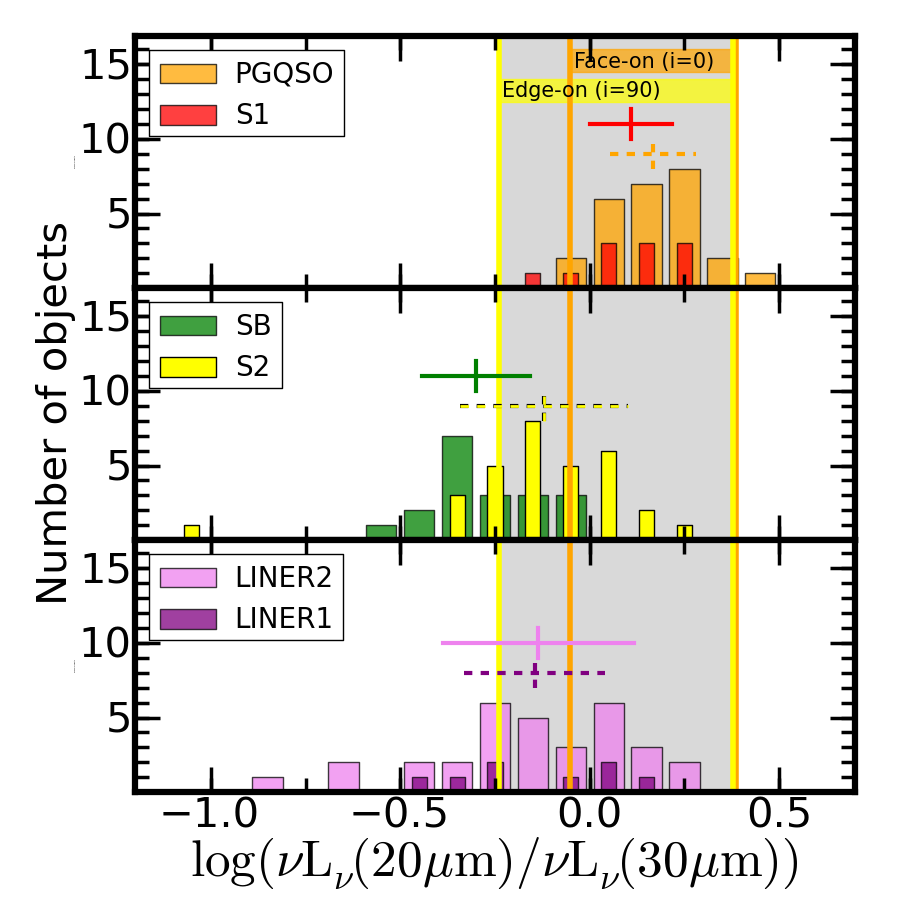}
\caption{Histograms of $\rm{log(\nu L_{\nu}(20\mu m)/\nu L_{\nu}(30\mu m))}$ for PG\,QSO (top-panel, orange-broad bars), S1 (top-panel, red-narrow bars), SB (middle-panel, green-broad bars), S2 (middle-panel, yellow-narrow bars), LINER1 (bottom-panel, purple-narrow bars), and LINER2 (bottom-panel, pink-broad bars). The mean values and one standard deviation over the mean for each class of objects are shown with large crosses (with the same color-code than the histogram), continuous lines for  PG\,QSO, SB, and LINER2 while dashed lines for S1, S2, and LINER1. The grey area of the plot shows the range of values expected for AGN according to the models given by \citet{Nenkova08}. The orange and yellow vertical lines show the same ranges but for inclination angles of $\rm{i=0^\circ}$ and $\rm{i=90^\circ}$, assuming that these values are representative of face-on and a edge-on AGN (see text). }
\label{fig:STEEPhisto}
\end{center}
\end{figure}

\begin{figure*}[!t]
\begin{center}
\includegraphics[width=2.\columnwidth]{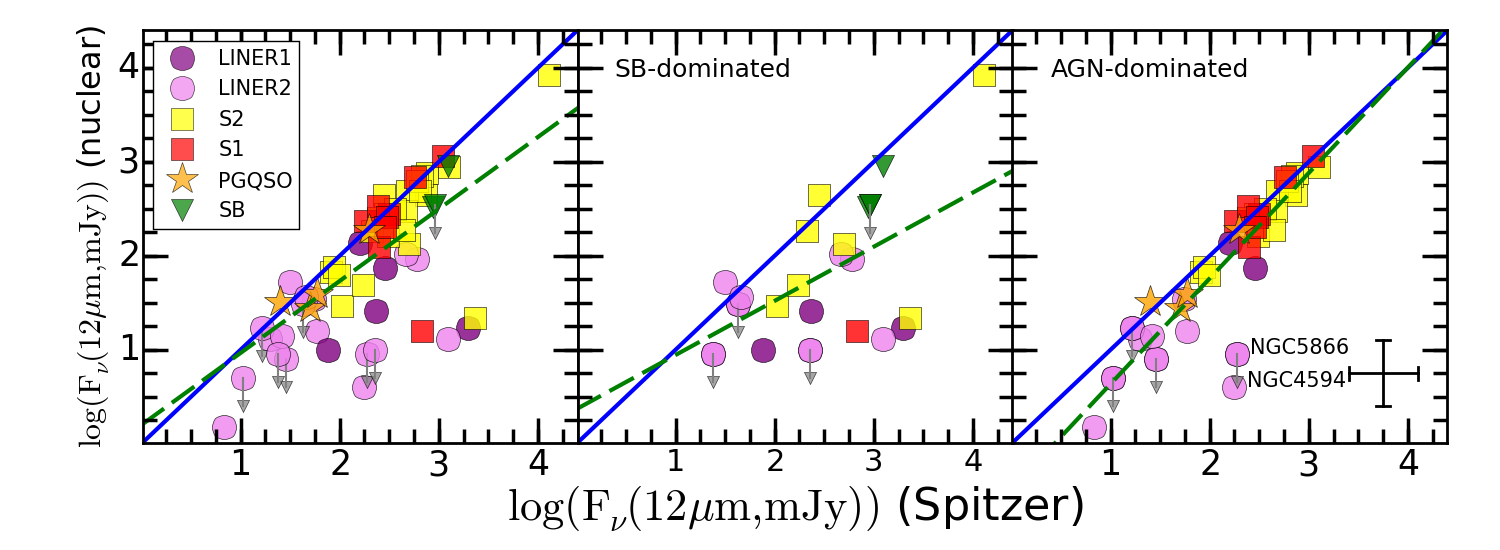}
\caption{The 12$\rm{\mu}$m flux obtained with ground-based telescopes (nuclear) versus 12$\rm{\mu}$m flux obtained with the \emph{Spitzer}/IRS spectra. Note that both quantities are shown in log-scales. A cross is shown to illustrate the error bars in these measurements (see Section \ref{sec:data}). The left panel shows the results for the entire sample of objects with ground-based telescopes, the middle panel shows the objects classified as SB-dominated and the right panel shows only objects with \emph{Spitzer}/IRS spectra classified as AGN-dominated in this work (see text). The blue continuous line represents the one-to-one relation and the green dashed line the best linear relation.}
\label{fig:LMIRcomparison}
\end{center}
\end{figure*}

\begin{itemize}
\item {Deep-Silicate}: Strength of the silicate feature above the maximum explained by clumpy models (i.e. $\rm{\tau_{9.7}>1.25}$) regardless of EW(PAH 6.2$\rm{\mu m )}$.  
\item {Strong-PAH}: $\rm{\tau_{9.7}<1.25}$ and EW(PAH 6.2$\rm{\mu m )>0.233\mu m}$.  
\item {Weak-PAH}: $\rm{\tau_{9.7}<1.25}$ and EW(PAH 6.2$\rm{\mu m )<0.233\mu m}$. 
\end{itemize}

This classification is included in Table \ref{tab:sample}. Starburst are located mostly in the region of strong-PAH. Two of them, though, populate the area of deep-silicates.
PG\,QSOs and S1s are placed in the region of weak-PAHs\footnote{The only exception is NGC\,5033 which is located in the region of the diagram of strong-PAHs.}. 

S2s tend to show larger $\rm{\tau_{9.7}}$ than S1s, as expected under the unified model of AGN. Similarly to S1s and PG\,QSOs, most S2s are in the region of weak-PAH. Only two of them are within the area of deep-silicates 
and five of them are within the area of strong-PAHs. 
Thus, according to this diagram, only seven out of the 31 S2s show signs of host-galaxy contamination at mid-IR.

LINERs mostly populate the area of weak-PAHs. All the LINER1s but NGC\,1097 are in this area of the diagram. Among LINER2s, two are in the region of deep-silicates 
and nine are in the strong-PAHs area. 
If LINERs in the weak-PAH area of the diagram are considered as AGN-dominated at mid-IR, then 30 out of the 41 are AGN-dominated at mid-IR.

\subsection{Steepness of the mid-IR spectra} 

The steepness of the mid-IR spectra characterises the relative contribution of warm and cool dust to the mid-IR \citep{Baum10}. Note that we refer here to the steepness at the spatial resolutions of the \emph{Spitzer}/IRS spectra (i.e. kpc scales) while at smaller scales (below 100 pc scales obtained with ground-based instruments), the mid-IR emission is expected to be dominated from dust heated by the AGN \citep{Honig11,Ramos-Almeida11}. It has been proven as a good indicator of the starburst content given its correlation with EW(PAH 11.3$\rm{\mu m}$)  \citep[e.g.][]{Wu09,Weedman05,Brandl06,LaMassa12}. This steepness has been defined in several bands by different authors; e.g. 20-30$\rm{\mu m}$ \citep{Baum10,Weedman05} or 15-30$\rm{\mu m}$ \citep{Brandl06,Wu09,Nardini08}. 
We present the 20 and 30$\rm{\mu m}$ luminosities to produce an estimate of the steepness of the mid-IR spectra in our sample. 

Fig. \ref{fig:STEEPhisto} shows the histogram of such steepness, expressed as $\rm{log(\nu L_{\nu}(20\mu m)/\nu L_{\nu}(30\mu m))}$ (see also Table \ref{tab:sample}). PG\,QSOs and S1s tend to show larger values of $\rm{log(\nu L_{\nu}(20\mu m)/\nu L_{\nu}(30\mu m))}$ than SBs. However, these two distributions overlap in the range $\rm{-0.25<log(\nu L_{\nu}(20\mu m)/\nu L_{\nu}(30\mu m))<0}$. S2s and LINERs show a wide range of $\rm{log(\nu L_{\nu}(20\mu m)/\nu L_{\nu}(30\mu m))}$, overlapping with S1s, PG\,QSOs, and SBs, although S1s and PG\,QSOs distributions are skewed toward the larger values of $\rm{log(\nu L_{\nu}(20\mu m)/\nu L_{\nu}(30\mu m))}$. Therefore, this ratio itself is not as good tracer of AGN dominance as it is EW(PAH 6.2$\rm{\mu m}$ ) (see previous subsection). Although in theory it is a good tracer of the contribution of warm and cool dust to the mid-IR, in practice, some SB-dominated spectra can have a $\rm{log(\nu L_{\nu}(20\mu m)/\nu L_{\nu}(30\mu m))}$ ratio consistent with those of AGN. 

Following the same idea than in the previous subsection, we have computed $\rm{log(\nu L_{\nu}(20\mu m)/\nu L_{\nu}(30\mu m))}$ of theoretical models obtained with the Clumpy libraries. The vertical-thick orange and yellow areas (Fig. \ref{fig:STEEPhisto}) show the range of steepness expected from these models for face-on (assuming $i=0$) and edge-on (assuming $i=90$) torii, respectively (delimited also with vertical orange and yellow, respectively). The expected range of values found for the models of Type-1 AGN is almost identical to the range of values found for PG\,QSOs and S1s. 
Moreover, the steepness of the spectra in the model of Type-2 AGN is expected to include the same range of values than S1s but extended toward lower values. Thus, Type-2 AGN are expected to show steeper spectra than Type-1 AGN. This is in agreement with our results, where S2s tend to show lower $\rm{log(\nu L_{\nu}(20\mu m)/\nu L_{\nu}(30\mu m))}$ than S1s. Seven out of the 20 SBs have a steepness fully consistent with the model of Type-2 AGN.
All the objects below the minimum steepness predicted for Type-2 AGN by the Clumpy libraries can be considered as SB-dominated ($\rm{log(\nu L_{\nu}(20\mu m)/\nu L_{\nu}(30\mu m))=-0.24}$). However, there are SBs above that limit. Thus, $\rm{log(\nu L_{\nu}(20\mu m)/\nu L_{\nu}(30\mu m))<-0.24}$ indicates that the spectrum is SB-dominated but we cannot discard that spectra showing $\rm{log(\nu L_{\nu}(20\mu m)/\nu L_{\nu}(30\mu m))>-0.24}$ might also be SB-dominated. 

We have classified as SB-dominated spectra those showing $\rm{log(\nu L_{\nu}(20\mu m)/\nu L_{\nu}(30\mu m))<-0.24}$. Among the S2s, six objects are therefore SB-dominated, 
three of them already classified as SB-dominated according to the strengths of the PAHs and silicate features. 
Combining both methods together, ten out of the 31 S2s (32\%) are SB-dominated. 

Thirteen out of the 41 LINERs show $\rm{log(\nu L_{\nu}(20\mu m)/\nu L_{\nu}(30\mu m))<-0.24}$. Two are LINER1s 
and the remaining 11 are LINER2s. Among them, eight were already classified as SB-dominated using the strengths of PAHs and silicate features.  
Interestingly, only two LINERs 
classified as SB-dominated with the diagram seen in Fig. \ref{fig:PAH62vsTau} are not SB-dominated using the steepness of the spectra. 
Fifteen out of the 40 LINERs (37.5\%) show signatures of being SB-dominated once the two methods presented in this section are considered together. The fraction of SB-dominated LINERs is similar to that of S2s. 

\begin{figure*}[!t]
\begin{center}
\includegraphics[width=1.\columnwidth]{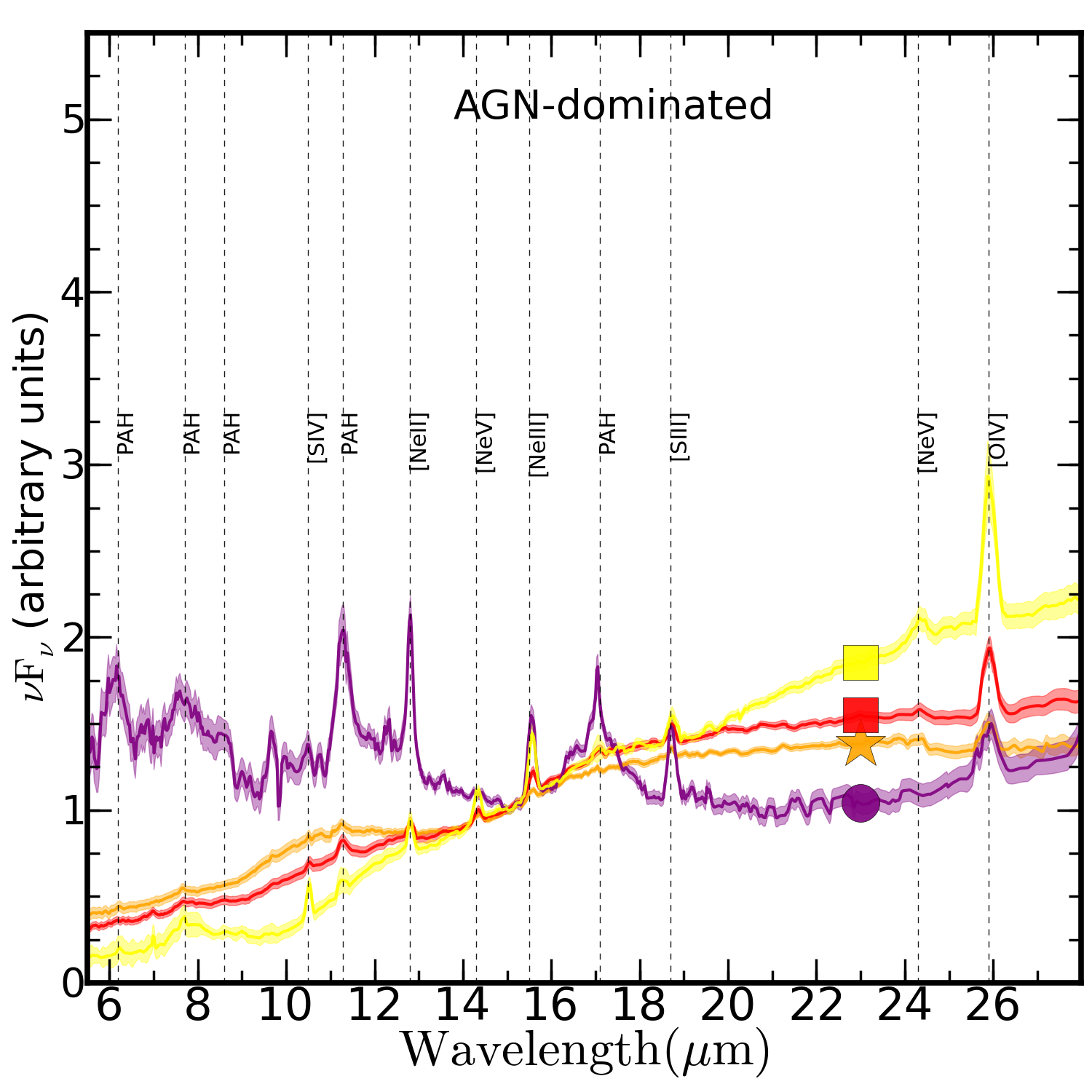}
\includegraphics[width=1.\columnwidth]{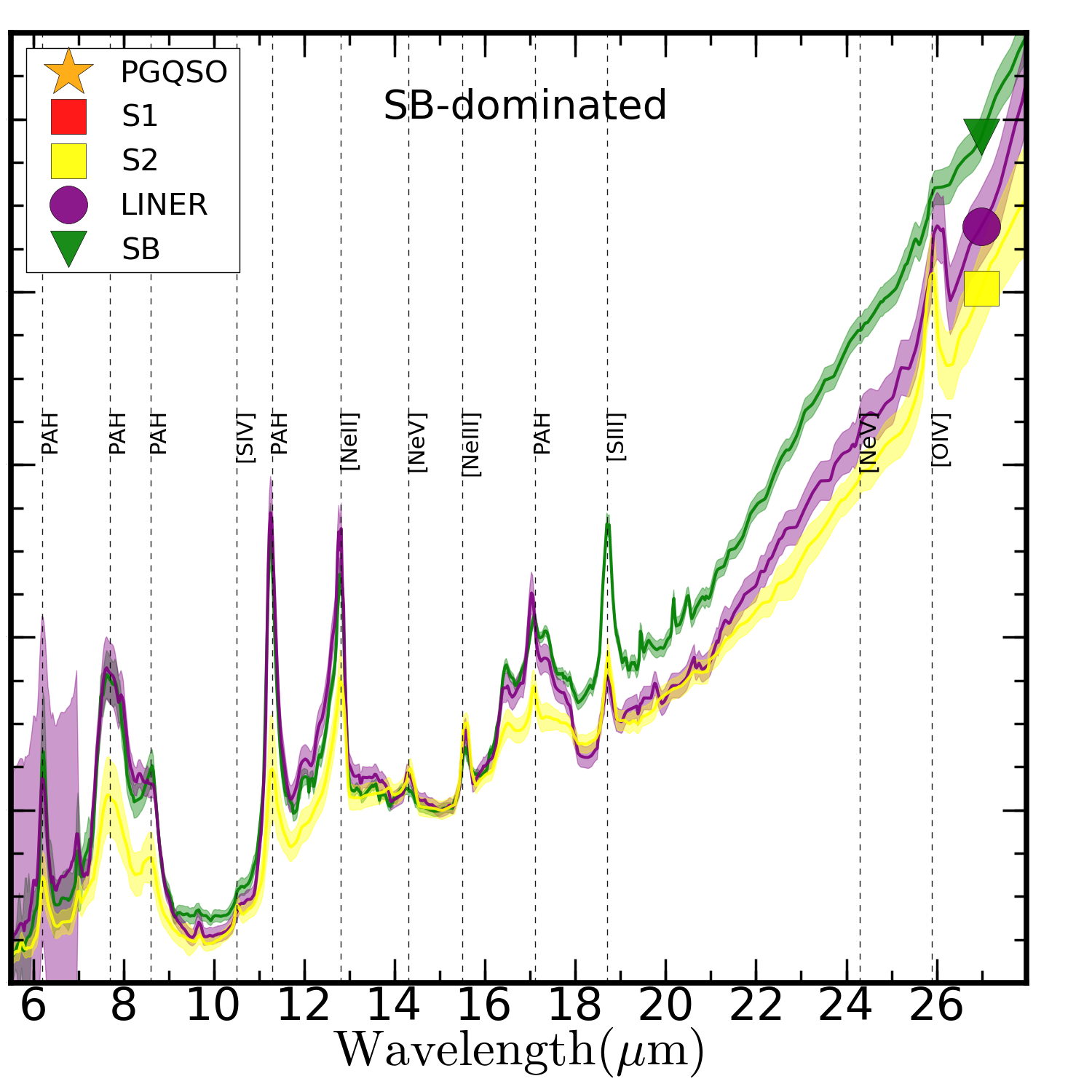}
\caption{(Left): average spectra for PG\,QSOs (orange), S1s (red), AGN-dominated S2s (yellow), and AGN-dominated LINERs (purple). (Right): average spectra for SBs (green), SB-dominated S2s (yellow), and SB-dominated LINERs (purple). We also show one standard deviation as a shaded region using the same colours. Each spectrum is also marked with different symbols at 27$\rm{\mu m}$ for clarity of the plot: PG\,QSOs (orange star), S1s (red square), SBs (green triangle), S2s (yellow square), and LINERs (purple circle). The average spectra are scaled to the flux at 15$\mu m$.}
\label{fig:SampleAGN}
\end{center}
\end{figure*}

\subsection{Goodness of the methodology to trace nuclear properties}

We have considered a mid-IR spectrum as AGN-dominated if it obeys these three criteria: EW(PAH 6.2$\rm{\mu m)<0.233 \mu m}$, $\rm{\tau_{9.7}<1.25}$, and $\rm{log(\nu L_{\nu}(20\mu m)/\nu L_{\nu}(30\mu m))>-0.24}$. To study the goodness of this method to select AGN-dominated \emph{Spitzer}/IRS spectra, we have plotted in Fig. \ref{fig:LMIRcomparison} the 12$\rm{\mu}$m flux for the \emph{Spitzer}/IRS spectra versus the same quantity for ground-based measurements for the 63 objects for which these measurements are available (see Section \ref{sec:data}). Ground-based and \emph{Spitzer}/IRS 12$\rm{\mu}$m fluxes show a linear relation although the dispersion is high (Pearson correlation coefficient of r=0.64, $\rm{P(null)=1.6\times10^{-7}}$). Moreover, the slope of the best linear fit (dashed line in the left panel of Fig. \ref{fig:LMIRcomparison}) is flatter than the one-to-one relation (continuous line). 

Most of the \emph{Spitzer}/IRS spectra with larger 12$\rm{\mu}$m flux than those from ground-based measurements are SB-dominated according to the EW(PAH $\rm{6.2 \mu m}$) method (see middle panel of Fig. \ref{fig:LMIRcomparison}). However, not all the SB-dominated \emph{Spitzer}/IRS objects show a 12$\rm{\mu}$m flux excess in the \emph{Spitzer}/IRS spectra compared to the ground-based measurements. We rule out the explanation of a distance effect in which more distant objects might include more SB-contribution in the nuclear spectra, because all our objects are nearby and no particular trend is found comparing SB- and AGN-dominated sources. Alternatively, this result might have two explanations: (1) our method to select SB-dominated spectra is too restrictive, and could include SB-dominated spectra that are actually AGN-dominated at 12$\rm{\mu}$m, and (2) the \emph{Spitzer}/IRS spectra do not contain extra emission compared to the ground-based measurements and both are tracing a nuclear SB-dominated spectrum. 

If we select only AGN-dominated \emph{Spitzer} spectra, the correlation between these two quantities improves (see Fig. \ref{fig:LMIRcomparison}, right panel) with a Pearson correlation coefficient of r=0.88 ($\rm{P(null)=5.7\times10^{-14}}$). The linear fit to the AGN-dominated sources (dashed line in Fig. \ref{fig:LMIRcomparison}, right panel) is very close to the one-to-one relation (continuous line in Fig. \ref{fig:LMIRcomparison}, right panel). The only two outliers are NGC\,4594 and NGC\,5866. Thus, when the AGN-dominated spectra are selected the nuclear 12$\rm{\mu}$m flux obtained with ground-based data is very close to the value obtained by the \emph{Spitzer}/IRS spectra. This reinforces our methodology as a good tool to isolate AGN-dominated mid-IR $\rm{Spitzer}$/IRS spectra. 


\begin{figure}[!t]
\begin{center}
\includegraphics[width=1.\columnwidth]{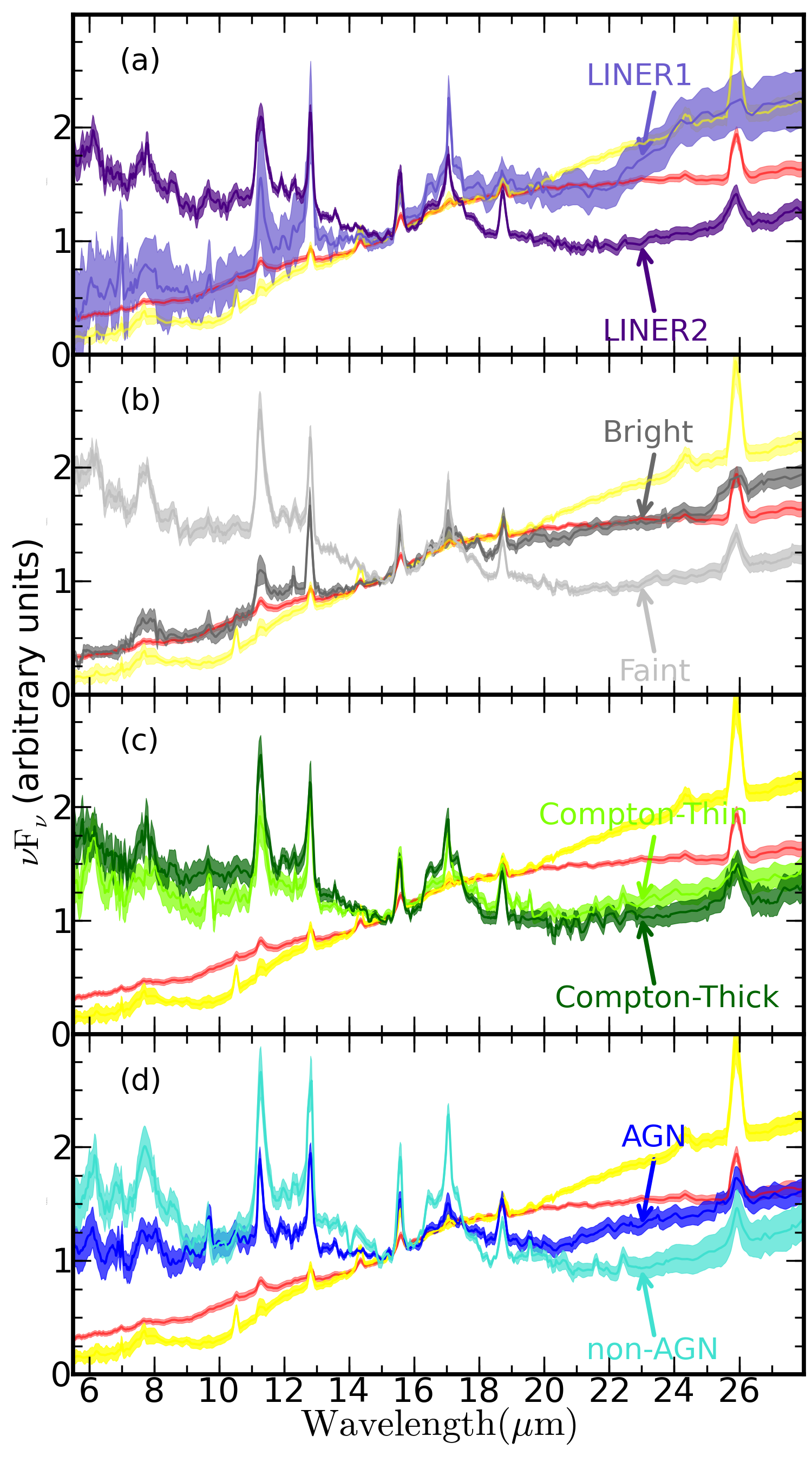}
\caption{The average spectra of AGN-dominated LINERs according to different subclassifications. From top to bottom: (a) LINER1s (light magenta) and LINER2s (dark magenta); (b) bright LINERs (dark grey) and faint LINERs (light grey); (c) Compton-thin (light green) and Compton-thick candidates (dark green); (d) objects classified at X-rays as AGN (dark blue) and non-AGN candidates (cyan). The average spectra for S1s (red) and AGN-dominated S2s (yellow) are also shown for comparison purposes.}
\label{fig:MeanLINERs}
\end{center}
\end{figure}

\section{Average spectra}\label{sec:AverageSpectra}

Fig. \ref{fig:SampleAGN} shows the average spectrum for each class of objects. These average spectra have been computed after normalising them to the flux at 15$\rm{\mu m}$. The shaded regions show the standard deviation over the average spectrum. We have computed the mean value for S2s and LINERs according to our mid-IR classification (see previous section) as AGN-dominated and SB-dominated (left and right panels, respectively). 

The average spectra of S1s and PG\,QSOs show similar shapes, showing the silicate feature in emission and similar steepness of the spectra (see left panel in Fig. \ref{fig:SampleAGN}). The relative differences between these two classes are an enhancement of the strength of the silicate feature in emission for PG\,QSOs compared to S1s and a slightly steeper spectrum for S1s compared to PG\,QSOs. The average SB spectrum is very different to that of S1s or PG\,QSOs (see right panel of Fig. \ref{fig:SampleAGN}). The main differences are strong PAH features, a steep spectrum, and deep silicate absorption features. Moreover, the classical lines associated with AGN emission such as [Ne V] at 14.3$\rm{\mu m}$ and 24.3$\rm{\mu m}$ or [O IV] at 25.9$\rm{\mu m}$, are clearly detected in the average spectra of PG\,QSOs and S1s but are absent in the average SB spectrum. Note here that the association of these lines with AGN emission have been questioned, finding them in some SB galaxies \citep[see][]{Pereira-Santaella10}. 

The average SB-dominated spectrum for S2s and LINERs (right panel of Fig. \ref{fig:SampleAGN}) are very similar to that of the SBs. The average AGN-dominated (left panel in Fig. \ref{fig:SampleAGN}) and SB-dominated (right panel in Fig. \ref{fig:SampleAGN}) spectra for S2s and LINERs are clearly different. This support our method as a good diagnostic of SB-dominated mid-IR spectra. This was also suggested by \citet{Alonso-Herrero14}, finding that the \emph{Spitzer} spectra of S1s and S2s are similar only if spectra with deep absorption silicate features are excluded from the analysis. The [Ne V] at 14.3$\rm{\mu m}$ and [O IV] at 25.9$\rm{\mu m}$ emission lines are clearly seen in the average spectra of SB-dominated S2s and LINERs. This was already shown by \citet{Dudik09}, finding these lines in a large fraction of LINERs. Thus, the average SB-dominated spectra of LINERs and S2s might still show mid-IR signatures of AGN nature, although the overall mid-IR spectra is not dominated by the AGN. 

The average S2 AGN-dominated spectrum (yellow spectrum in the left panel of Fig. \ref{fig:SampleAGN}) does not mimic S1s or PG\,QSOs. This average spectrum is steeper than those of S1s or PG\,QSOs. It also shows the silicate features in absorption while S1s and PG\,QSOs show an average spectra with silicate features in emission. This is expected since the silicate feature at $\rm{9.7\mu m}$ and $\rm{18\mu m}$ are predicted to be in emission for Type-1 AGN and in absorption for Type-2 AGN \citep{Nenkova08}. These predictions have already been confirmed by observations \citep[e.g.][]{Shi06}. On the similarities, the average (AGN-dominated) S2 spectrum shows the presence of [Ne V] at 14.3 and 24.3$\rm{\mu m}$ and [O IV] at 25.9$\rm{\mu m}$ emission lines. 

The average AGN-dominated LINER spectrum (purple spectrum in the left panel of Fig. \ref{fig:SampleAGN}) can be clearly distinguished from PG\,QSOs, S1s, S2s, and SBs. The main characteristic of this average spectrum is the rather flat continuum all over the 6-28 $\rm{\mu m}$ wavelength range. Moreover, it shows strong PAH features at 11.3$\rm{\mu m}$ and 17$\rm{\mu m}$. The [O IV] at 25.9$\rm{\mu m}$ emission line is prominent as in S1s, PG\,QSOs, and S2s. However, the [Ne V] at 14.3$\rm{\mu m}$ emission line is clearly undetected as is the [Ne V] at 24.3$\rm{\mu m}$. Below 20$\rm{\mu m}$ this spectrum resembles that of SBs. However, it can be clearly distinguished from SBs because the average spectrum of LINERs do not show a steep spectrum and it lacks of the silicate absorption features seen in SBs. Moreover, LINERs also show the [O IV] at 25.9$\rm{\mu m}$ that the SBs do not show. 

\begin{figure*}[!t]
\begin{center}
\includegraphics[width=2.\columnwidth]{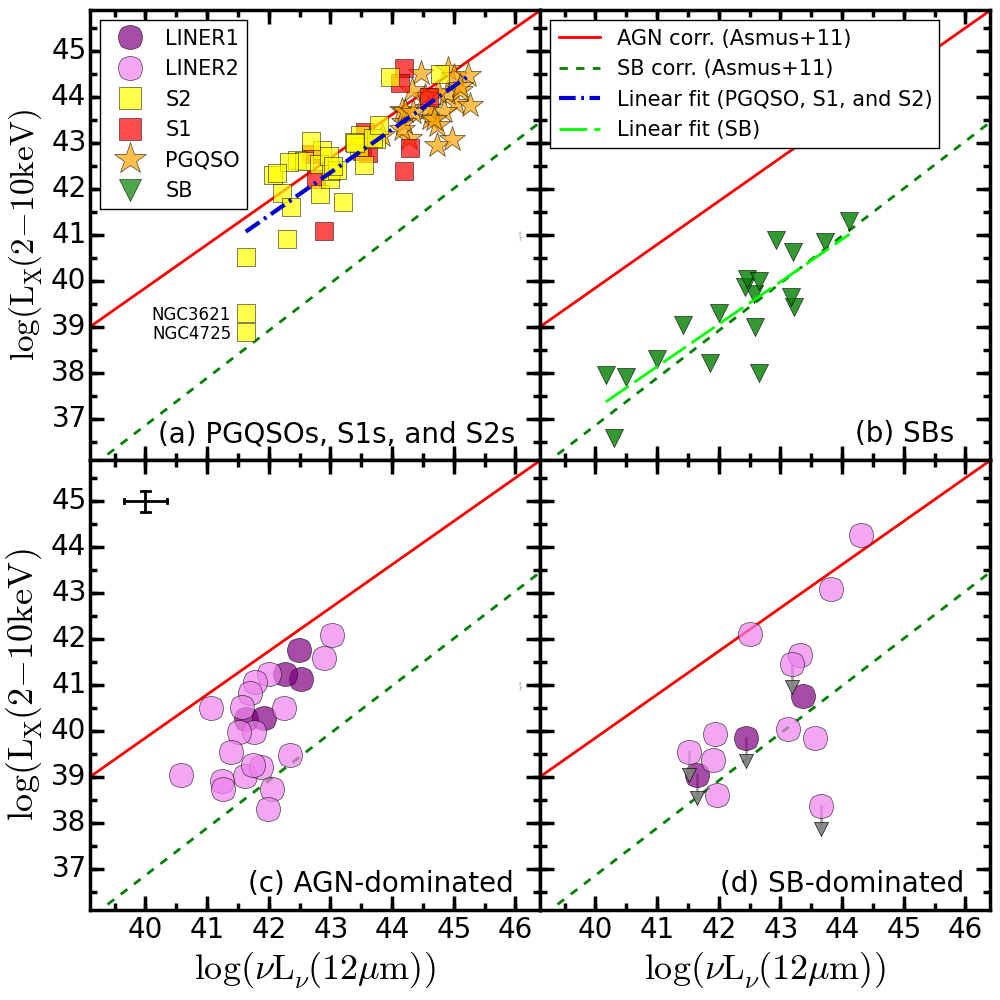}
\caption{The 2-10 keV luminosity versus the 12$\rm{\mu}$m luminosity, both in logarithmic scale for PG\,QSOs, S1s and S2s (a), SBs (b), AGN-dominated LINERs (c) and SB-dominated LINERs (d). The continuous-red and dashed-green lines show the best fit correlations for AGN and SBs, respectively, reported by \citet{Asmus11}. The typical error for these measurements is shown as a cross in the top-left corner of panel (c). Errors for the X-ray luminosity are estimated as 10\% of its value. The dot-dashed blue line and long-dashed light green line show the linear fit for PG\,QSOs, S1s, and S2s and for SBs, respectively. Grey arrows mark objects with reported upper-limits on the X-ray luminosity. }
\label{fig:LXvsMIR}
\end{center}
\end{figure*}

\begin{figure*}[!t]
\begin{center}
\includegraphics[width=2.\columnwidth]{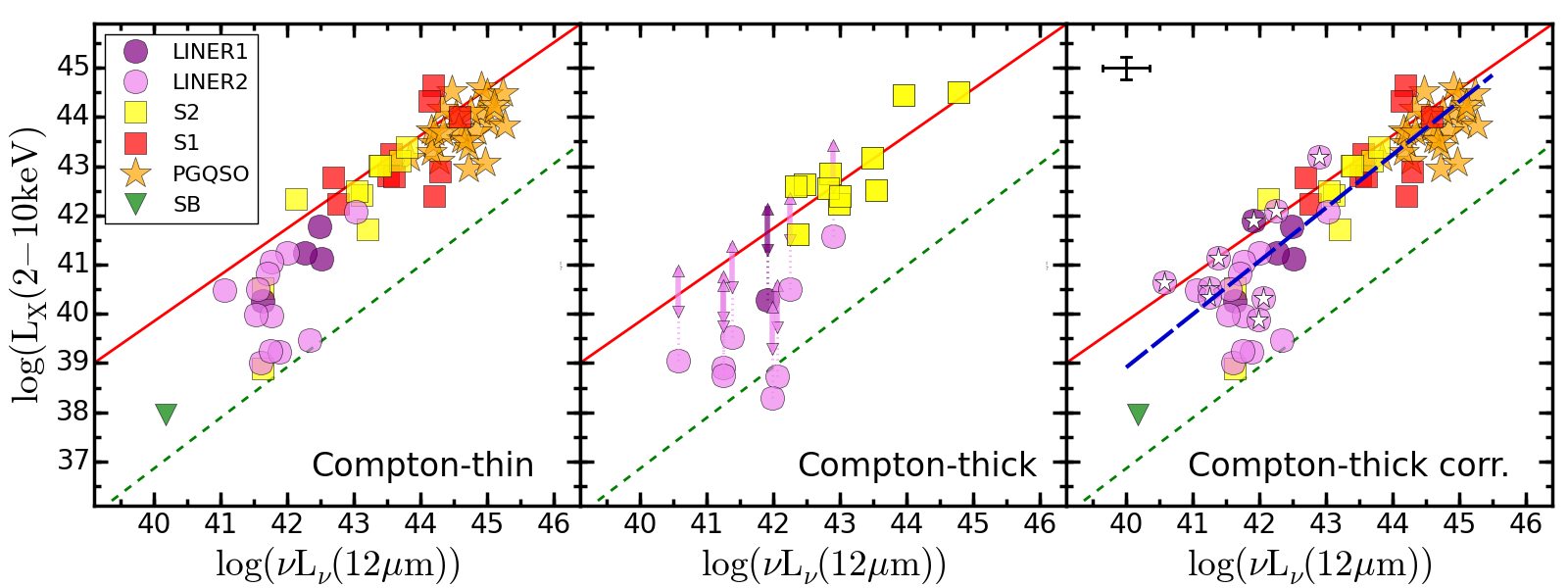}
\caption{The 2-10 keV luminosity versus the 12$\rm{\mu}$m luminosity, both in logarithmic scales for Compton-thin (left), Compton-thick (middle), and the full sample after Compton-thickness correction (right). The typical error for these measurements is shown as a cross in the top-left corner of the right panel. The continuous-red and dashed-green lines show the best fit correlations for AGN and SBs, respectively, reported by \citet{Asmus11}. The long-dashed blue line shows the linear fit to the full sample. Grey arrows mark objects with reported upper-limits on the X-ray luminosity. Small double arrows in the middle panel show the expected locus if the real intrinsic 2-10 keV luminosity is 10-70 times higher than that estimated. }
\label{fig:LXvsMIRCthickness}
\end{center}
\end{figure*}

\subsection{Sub-classes of LINERs}\label{sec:Subclasses}

As explained in the introduction, the LINERs are a heterogenous family of objects. In order to study the subclasses of LINERs, Fig. \ref{fig:MeanLINERs} shows the average spectra for several subclassifications of LINERs. Note that these average spectra have been computed including only AGN-dominated spectra. 

The average spectrum of objects optically classified as LINER1s (panel (a), light magenta spectrum in Fig. \ref{fig:MeanLINERs}) is steeper than that of LINER2s (panel (a), dark magenta spectrum). In fact, the average spectrum of Type-1 LINERs is consistent with that of AGN-dominated S2s (yellow spectrum). However, the dispersion in the average spectrum of LINER1s is quite large because, among the five AGN-dominated LINERs1, NGC\,4450 shows a flat spectrum which is not consistent with the average trend for this class.

We have also classified the AGN dominated LINERs into two classes attending to their $\rm{L_{X}(2-10~keV)}$: bright LINERs for objects with $\rm{L_{X}(2-10~keV)>10^{41}erg~s^{-1}}$ and faint LINERs for those with X-ray luminosities below that limit. This is the limit where the torus is expected to disappear at a bolometric luminosity of $\rm{L_{bol}\sim10^{42}erg~s^{-1}}$\citep[][]{Elitzur06}\footnote{This assumes a conversion between the X-ray luminosity and the bolometric luminosity of $\rm{L_{bol}\simeq10\times L_{X}(2-10~keV)}$ \citep{Ho08}. This conversion factor is the most conservative value we have found \citep{Ho09}. Note that any higher value could include more objects as bright LINERs (see also the discussion).}. Among the AGN-dominated LINERs, 7 are classified as bright LINERs and 18 are classified as faint LINERs. The resulting average spectra of bright and faint LINERs (dark and light grey spectra in panel (b) of Fig. \ref{fig:MeanLINERs}) are quite similar to that of LINER1s and LINER2s. Thus, bright LINERs show a steeper spectrum compared to faint LINERs, compatible with S1s. Note that the dispersion of the average spectrum of bright LINERs is much lower than that of LINER1s. We have investigated if a different morphology of the host galaxy for bright and faint LINERs could produce the differences in their average mid-infrared spectra. Interestingly, NGC\,4450 is the only AGN-dominated Type-1 LINER hosted in a late-type galaxy. Only four (out of the 18) faint LINERs and 1 (out of the 7) bright LINERs are hosted in late type galaxies (i.e. t$\rm{>}$1). This lack of late type galaxies hosting LINERs is expected since they are usually hosted in early type galaxies \citep{Carrillo99}. Thus, we have not found any particular tendency in their host-galaxy morphology. Indeed, the average spectrum of faint LINERs hosted in late type galaxies is consistent with the average spectrum of faint LINERs (and with the average spectrum of faint LINERs hosted in early type galaxies).

Dark and light green spectra shown in panel (c) of Fig. \ref{fig:MeanLINERs} show the average spectra for Compton-thin and Compton-thick candidates reported by \citet{Gonzalez-Martin09B}. Compton-thick sources are expected to be so obscured ($\rm{N_{H}>1.5\times 10^{25}cm^{-2}}$) that the bulk of the X-ray continuum emission of the AGN is only seen above 10 keV. Sources were classified according to the ratio between the fluxes of [O III]$\rm{\lambda 5007\AA}$ and the 2-10 keV fluxes, the EW of the FeK$\rm{\alpha}$ emission line, and the hard X-ray slope of the spectrum. Sixteen of them were classified as Compton-thin LINERs and nine were classified as Compton-thick candidates. We do not find significant differences between the average mid-IR spectra of Compton-thin and Compton-thick candidates. 


\citet{Gonzalez-Martin09A} classified 22 out of the 25 AGN-dominated LINERs at X-rays: 16 AGN candidates and 6 non-AGN candidates. Both average spectra are quite flat (cyan and blue spectra in panel (d) of Fig. \ref{fig:MeanLINERs}). Interestingly, PAH features and [Ne II] at 12.8$\rm{\mu m}$ strengths are larger for non-AGN candidates than for AGN candidates suggesting the dominance of the host galaxy contribution.

\section{X-ray versus mid-IR luminosities}\label{sec:XrayMIR}

In AGN, the X-ray emission is thought to originate in the innermost regions of the accretion flow by Comptonisation processes. The mid-IR emission is generally dominated by thermal emission by dust at parsec-scale distances from the SMBH. The tight correlation found between the X-ray and mid-IR emission of AGN supports this physical connection, regardless of the Seyfert type \citep[][]{Krabbe01,Gandhi09,Levenson09}. Furthermore, SB nuclei also show a relation between X-ray and mid-IR luminosities but it is offset when compared to that of AGN \citep[][]{Asmus11}. The study of this correlation could bring light on the dominant process involved in LINERs.  

Fig. \ref{fig:LXvsMIR} shows $\rm{L_{X}(2-10~keV)}$ versus $\rm{\nu L_{\nu}(12\mu m)}$ for our sample. Note that $\rm{\nu L_{\nu}(12\mu m)}$ is computed using the \emph{Spitzer} spectra. S1s, S2s, and PG\,QSOs (panel (a)) follow the same relation previously found for AGN \citep[continuous red line, from][]{Asmus11}. The Pearson correlation coefficient for them is $\rm{r = 0.82}$ ($\rm{P(null)=4.7\times 10^{-18}}$) and the best linear fit (blue dot-dashed line) is very close to that reported by \citet{Asmus11} using high spatial resolution mid-IR images. Therefore, this correlation holds for most S1s, S2s, and PG\,QSOs even with the relatively low spatial resolution of the \emph{Spitzer}/IRS spectra. This may suggest that for these sources the AGN continuum dominates the \emph{Spitzer}/IRS emission. Only two S2s (NGC\,3621 and NGC\,4725) fall close to the correlation found for SB nuclei \citep[dashed green line, also from][]{Asmus11}. In our study NGC\,3621 was classified as SB-dominated and NGC\,4725 as AGN-dominated (see Section \ref{sec:AGNvsSB}). SB nuclei (green up-side down triangles in the panel (b) in Fig. \ref{fig:LXvsMIR}) also fall into the expected correlation for them with a correlation coefficient $\rm{r=0.83}$ ($\rm{P(null)=4.5\times 10^{-11}}$). None of them are compatible with the AGN correlation. Thus, this correlation seems to be very effective for distinguishing pure AGN emission from SB emission. 

The bottom panels of Fig. \ref{fig:LXvsMIR} show $\rm{L_{X}(2-10~keV)}$ versus $\rm{\nu L_{\nu}(12\mu m)}$ for LINERs. We split the plot into the AGN-dominated (panel (c)) and the SB-dominated (panel (d)) LINERs. Most LINER nuclei are placed between the AGN and the SB linear relations. The linear correlation for them is not significant ($\rm{r=0.68}$ and $\rm{P(null)=8\times 10^{-6}}$).  Low luminosity LINERs are near SB correlation and high luminosity LINERs are parallel to the AGN correlation. This could be due to an underestimation of the X-ray luminosity (see below). 

Some LINERs behave as AGN while some others are similar to SBs (see Section \ref{sec:AGNvsSB}). Many of the LINER nuclei classified as SB-dominated are placed along the SB correlation (all the Type-1 LINERs). Again, this confirms our method described in Section \ref{sec:AGNvsSB} as a good tool to discriminate SB- from AGN-dominated spectra. However, 
three LINER2s (NGC\,3079, NGC\,6240, and NGC\,7130) are placed in the AGN correlation but were classified as SB-dominated in the mid-IR. Interestingly, all of them are known Compton-thick AGN. This suggests that the 12$\rm{\mu m}$ flux could be AGN-dominated although the full \emph{Spitzer}/IRS spectrum is SB-dominated. Therefore, irrespective of the mid-IR spectral shape, the 12$\rm{\mu m}$ flux is a good tracer of the AGN power \citep[e.g.][]{Gonzalez-Martin13}. 

The inclusion of LINERs in this correlation allows to validate it at lower luminosities \citep[$\rm{L_{X}(2-10~keV})<10^{42}erg~s^{-1}$, see also][]{Mason12}. The final linear correlation for AGN-dominated spectra is: 

\begin{eqnarray}
log(L_X) = (-12.34\pm0.05) + (1.26\pm0.01)log(\nu L_{\nu}(12\mu m))
\end{eqnarray}

\noindent which is very significant ($\rm{r=0.92}$ and $\rm{P(null)=2\times 10^{-28}}$). However, the slope of this correlation is steeper ($\rm{\sim 1.26}$) than that previously found \citep[$\rm{\sim}$1.06, see][]{Asmus11}. This excess in mid-IR luminosity for faint LINERs was already found by \citet{Mason12}. They argued that this discrepancy could be due to optically thin material that obscures the inner parts of the AGN because most of them showed silicate features in emission. However, our faint LINERs do not show in average silicate features in emission (see Section \ref{sec:Subclasses} and Fig.~\ref{fig:MeanLINERs}). 

The most natural explanation is that the Compton-thick nature of some low-luminosity AGN results in an underestimation of the true $\rm{L_{X}(2-10~keV)}$ of these sources. These $\rm{L_{X}(2-10~keV)}$ estimates come from studies using the spectra of LINERs at energies below 10 keV. However, Compton-thick sources show the bulk of the AGN power at energies above 10 keV. The intrinsic luminosity could be 10-70 times higher than the estimated using only energies below 10 keV in the Compton-thick scenario \citep{Maiolino98}. \citet{Gonzalez-Martin09B} classified around 50\% of their LINER sample as Compton-thick candidates. Fig. \ref{fig:LXvsMIRCthickness} shows the AGN-dominated objects in our study attending to their Compton-thin (left) and Compton-thick (middle) classification. Most of the Compton-thin sources are close to the previously reported correlation for AGN. Compton-thick S2s are nicely placed along the AGN correlation found by \citet{Asmus11}. This is expected because the X-ray luminosities included for S2s are all corrected for their Compton-thick nature \citep[most of them included in][]{Goulding12}, either using X-ray measurements above 10 keV or assuming a factor between the observed and intrinsic X-ray luminosity for other Compton-thick AGN \citep[see][for details in this conversion factor]{Panessa06}. Compton-thick LINER candidates tend to be shifted toward X-ray luminosities lower than predicted for the AGN correlation. Note, however, that most of them are not consistent with the SB correlation either. The double arrows of Fig.~\ref{fig:LXvsMIRCthickness} (middle panel) show their expected locus if the X-ray luminosity were $\rm{\sim}$10-70 times higher. Most of the Compton-thick LINERs can be placed in the AGN correlation if the correction is applied. The linear fit to the entire sample (long-dashed blue line in Fig. \ref{fig:LXvsMIRCthickness}), once the intrinsic X-ray luminosity is corrected results in $\rm{L_{X}(2-10~keV)(intrinsic)= 40\times L_{X}(2-10~keV)(observed)}$ for Compton-thick LINERs (marked as white stars in Fig. \ref{fig:LXvsMIRCthickness}, right panel), which is very close to the linear relation found for AGN. Nonetheless, four LINERs and one S2 still remain very close to the SB correlation. 

\section{Discussion}\label{sec:discussion}

The nature of LINER nuclei has been extensively studied since they were firstly discovered by \citet{Heckman80}. Using multiwavelength information and several techniques as compactness, hardness, and variability, we now know that a large fraction of them host an AGN \citep{Maoz05,Gonzalez-Martin09A,Gonzalez-Martin09B,Ho08}. However, what makes them a unique class is still unknown.  \citet{Singh13} showed that post-main sequence stars might be an important contributor to the optical frequencies. Interestingly, they argue that this population might be present in more powerful AGN, although it is outshined by the AGN itself. At X-rays, several authors have pointed out to obscuration as one of the main ingredients for their different nature\citep{Dudik09,Gonzalez-Martin09B}. Therefore, mid-IR frequencies are key to study this obscuration since the emission absorbed at optical and UV frequencies is expected to be reprocessed at those wavelengths. 

We have compiled a sample of 40 mid-IR spectra of LINERs observed with \emph{Spitzer}/IRS and have compared them to samples of SBs, S2s, S1s, and PG\,QSOs. Although the low spatial resolution of these \emph{Spitzer}/IRS spectra is a disadvantage, we have been able to isolate SB-dominated from those which are not by using well known mid-IR tools (see Sect. \ref{sec:AGNvsSB}). We focus this discussion in two main issues about the obscuration of LINERs in light of the present results: (1) Compton-thickness and (2) torus signatures in LINERs.

\subsection{Compton-thickness} \label{sec:compton-thickness}

\citet{Gonzalez-Martin09B} showed that a large fraction (up to 50\%) of their LINER sample might be Compton-thick, i.e. with such a high obscuration that the intrinsic continuum of the AGN is fully suppressed at energies below 10 keV. This was done using indirect arguments as the [O III]$\rm{\lambda 5007\AA}$ and to X-ray flux-ratio or the EW of the neutral FeK$\rm{\alpha}$ line at 6.4 keV. However, this is not yet confirmed with direct observations because LINERs are too faint to be observed above 10 keV with the open X-ray instrumentation to the community\footnote{NuSTAR satellite is able to observe these faint source but it is part of a close collaboration.}.

We have take advantage of using the X-ray to mid-IR correlation found for AGN to study the Compton-thick nature of LINERs. LINERs classified as Compton-thick candidates by \citet{Gonzalez-Martin09B} are systematically located below the relation found for AGN (see Fig. \ref{fig:LXvsMIRCthickness}, middle panel). This is naturally explained if they are indeed Compton-thick AGN, since their X-ray intrinsic luminosity is underestimated by a factor of 10 or more. This was already found by \citet{Mason12} for a small number of objects. Four additional LINERs and one Seyfert also have lower X-ray luminosities that expected according to their mid-IR luminosities (see Fig. \ref{fig:LXvsMIRCthickness}, left panel). Thus, either they host a Compton-thick nucleus or are SB-dominated. This suggests that the actual fraction of Compton-thick LINERs may be even higher than previously inferred \citep{Gonzalez-Martin09B}. This large fraction of obscuration might explain why post-main sequence stars dominate the optical spectrum in LINERs \citep{Singh13}, while the AGN completely dominates the emission in powerful and/or less obscured AGN.

Observationally, a large fraction of AGN in the local Universe are obscured by Compton-thick gas \citep{Maiolino98,Matt00}. From the theoretical point of view, a sizeable population of mildly Compton-thick sources is postulated in all the AGN synthesis models for the X-ray background in order to match the intensity peak of the XRB spectrum at about 30 keV \citep{Comastri04}. These theoretical and observational evidence are consistent with the results of the present paper. Furthermore, the fraction of Compton-thick Seyferts is lower than that reported in LINERs (and confirmed in this analysis). This is fully consistent with the paradigm in which the fraction of obscured sources increases when the luminosities decrease in the high-redshift Universe \citep{Ueda01} and in the local Universe \citep{Shinozaki06}.

\begin{figure}[!t]
\begin{center}
\includegraphics[width=1.\columnwidth]{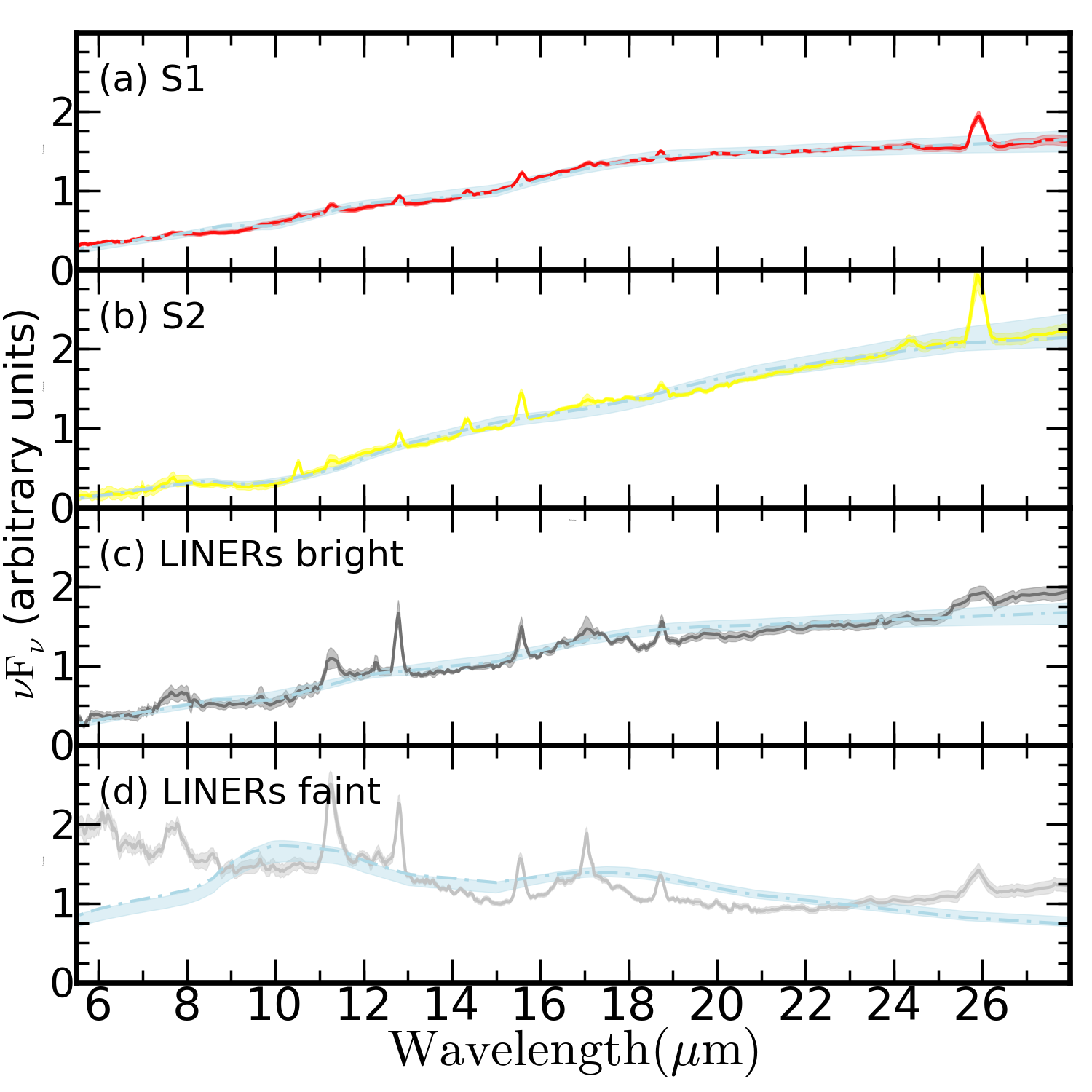}
\caption{Best-fit to clumpy models (dot-dashed light-blue line) for the average spectra, from top to bottom, of S1s (panel (a)), S2s (panel (b)), bright LINERs (panel (c)), and faint LINERs (panel (d)). Note that bright LINERs are those with an X-ray luminosity above $\rm{L_{X}>1\times 10^{41} erg/s}$. All the spectra are normalised to their flux at 15$\rm{\mu m}$. The shaded light-blue area shows the lower and upper bounds obtained with the clumpy models (see text). }
\label{fig:LINERsmodels}
\end{center}
\end{figure}

\begin{figure*}[!t]
\begin{center}
\includegraphics[width=2.\columnwidth]{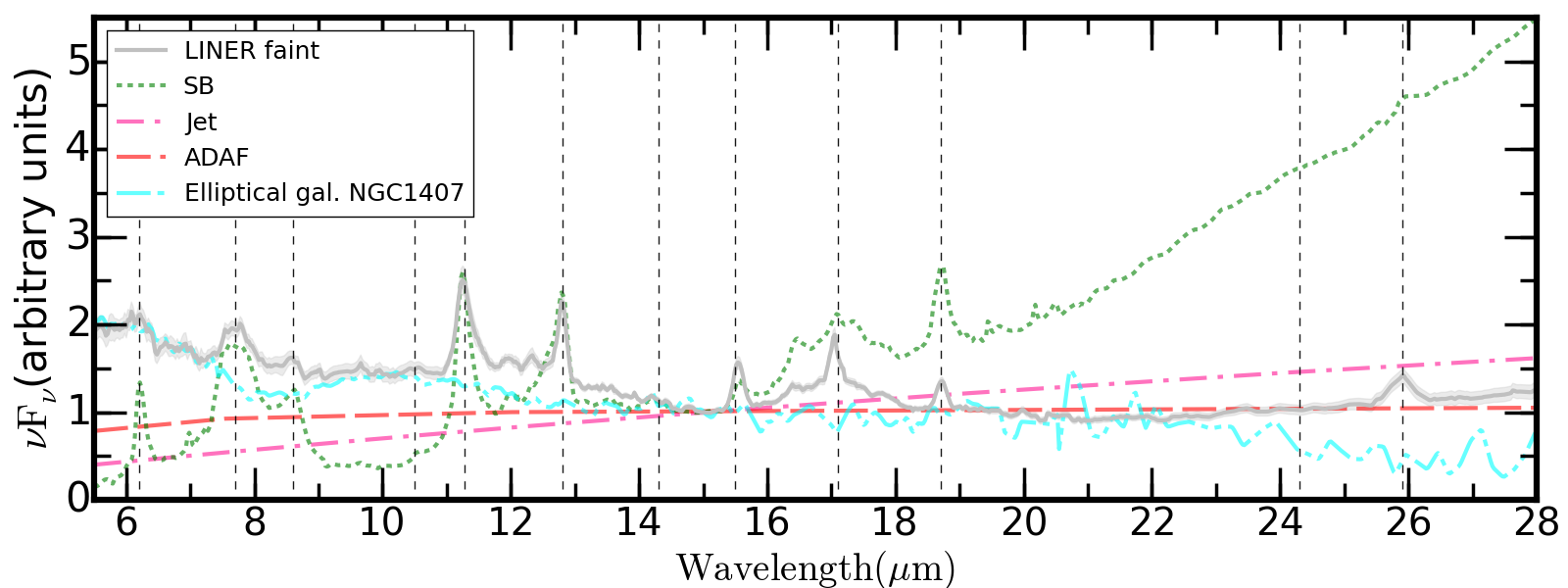}
\includegraphics[width=2.\columnwidth]{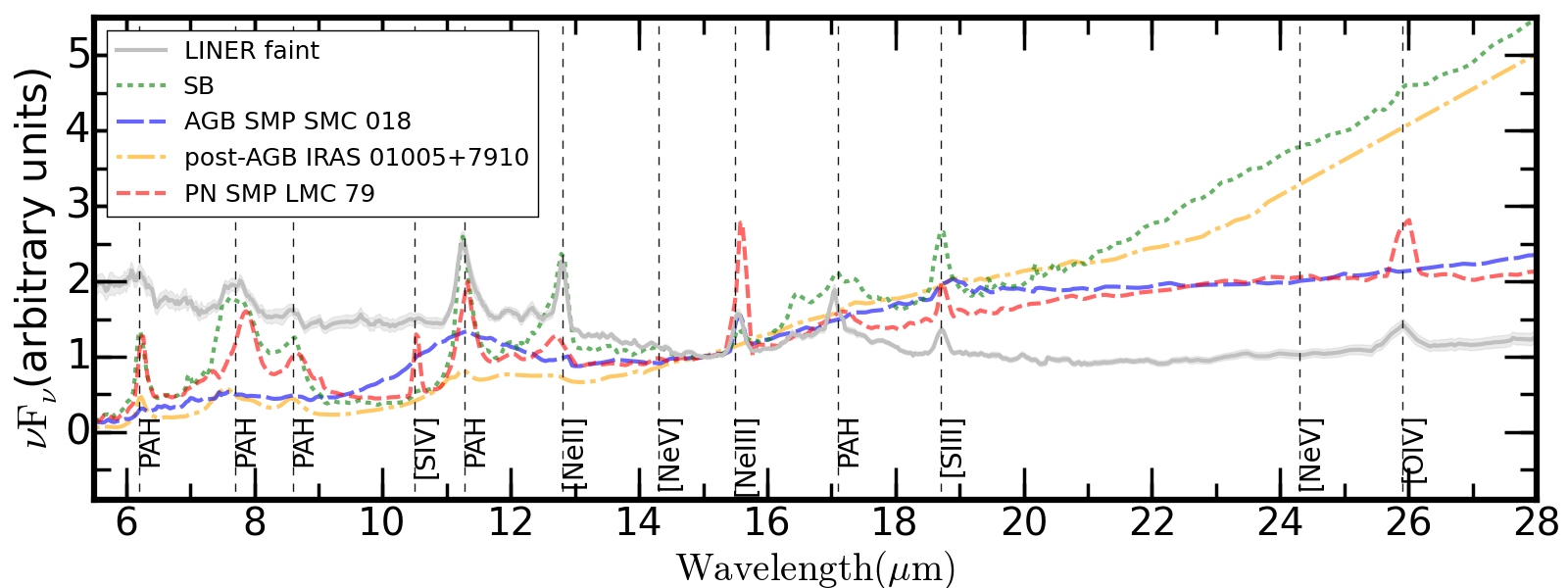}
\caption{(Top): Mid-IR spectrum of faint LINERs (grey continuous line), SBs (green dotted line), elliptical galaxy NGC\,1407 (cyan dot-dashed line), synchrotron emission of a jet (pink dot-dashed line), ADAF (red dashed line), and . (Bottom): Mid-IR spectrum of faint LINERs (grey continuous line), SBs (green dotted line), AGB star SMP\,SMC\,018 (long-dashed blue line), post-AGB star IRAS\,01005+7910 (dot-dashed orange line), and PN SMP\,LMC\,79 (short-dashed red line). }
\label{fig:JetElipAdaf}
\end{center}
\end{figure*}

\subsection{Torus in LINERs}\label{sec:torus}

We have found that the average mid-IR spectrum of LINERs is markedly different from those of SBs, S1s, S2s, or PG\,QSOs (see Fig. \ref{fig:SampleAGN}). The former shows a rather flat continuum from 6 until 28$\rm{\mu m}$. On top of this flat continuum, prominent emission lines of [Ne II], [Ne III], [S III], and [O IV] are clearly seen. The [O IV] emission line suggests the presence of AGN emission although it could be explained by another emission mechanisms \citep[see][]{Pereira-Santaella10}. The PAH features are also clearly seen. They indicate a non-negligible contribution of star-formation to the mid-IR spectra of LINERs. However, note that star-formation cannot be the dominant process for them because the SB mid-IR continuum is much steeper\footnote{Note that in this discussion SB-dominated LINERs are excluded, see Section \ref{sec:AGNvsSB}.} (see right panel of Fig. \ref{fig:SampleAGN}). 

We have also found that the shape of the average mid-IR spectrum of LINERs does not depend on its Compton-thickness or AGN nature at X-rays (see Fig. \ref{fig:MeanLINERs}). It depends on the optical type, i.e. the average mid-IR spectrum of LINER1s is different to that of LINER2s. However, the scatter around the mean in LINER1s are large indicating that probably they are not a well defined class. The best distinction between two subclasses is found for bright and faint LINERs, above and below $\rm{L_{X}>1\times 10^{41} erg/s}$, respectively (Fig. \ref{fig:MeanLINERs}). 

The mid-IR spectrum of AGN is expected to be dominated by dust re-emission of optical/UV emission. Under the unified model of AGN this dust is distributed in a dusty torus \citep{Antonucci93}. This torus was first postulated to be a smooth doughnut-like structure. However, nowadays the preferred scenario is a clumpy structure, i.e. made of dusty clouds within a toroidal distribution \citep{Krolik88,Pier92}. \citet{Nenkova02} developed a formalism for handling radiative transfer in clumpy media and applied it to the IR emission from the AGN dusty torus. \citet{Nenkova08} applied this formalism to develop a set of models for this clumpy medium. They showed that clumpy torus models are consistent with current AGN observations if they contain a number of dusty clouds along the equatorial axis of $\rm{N_{o}\sim 5- 15}$, each with an optical depth of $\rm{\tau_{V}\sim 30-100}$. \citet{Ramos-Almeida11} modelled the IR SED for a sample of S1s and S2s by \citet{Nenkova08} dusty torus models, using a Bayesian inference tool called BayesClumpy \citep{Asensio-Ramos09}. They found that the IR SED of both S1s and S2s are well fitted by these clumpy torus models although their intrinsic properties are different \citep[larger covering factor for S2s than for S1s, see also][]{Alonso-Herrero11}. 

We have used the BayesClumpy tool to investigate whether clumpy models could fit our average mid-IR spectra. 
The median best fit models found for the average mid-IR spectra are shown in Fig. \ref{fig:LINERsmodels} as dot-dashed light-blue line together with 68\% confidence intervals for all the parameters as light-blue filled regions \citep[see][for more details on the modelling]{Ramos-Almeida09}. As already shown by \citet{Ramos-Almeida11}, the continuum shape of S1s and S2s are very well represented by clumpy models (panels (a) and (b)). Bright LINERs (i.e. $\rm{L_{X}(2-10~keV)>1\times 10^{41} erg/s}$, panel (c)) can also be reproduced by these models. However, faint LINERs (panel (d)) are poorly represented by them. This may suggest that the optically thick torus emission may not longer dominate the mid-IR emission of faint LINERs. 

In the last years, several pieces of observational evidence have supported the scenario in which a single continuum distribution of clouds within a wind is responsible for both the BLR and the dusty torus \citep[][and references therein]{Elitzur06}. Under this scenario, the difference between the BLR and dusty torus is just a change on the composition of this wind at the dust sublimation radius. An immediate consequence of the disk-wind scenario is the prediction that the torus and the BLR disappear at bolometric luminosities below $\rm{L_{bol}\sim 10^{42} erg/s}$ \citep{Elitzur06}. This limit is consistent with our results assuming a conversion factor between the bolometric luminosity and X-ray luminosity of $\rm{L_{bol}/L_{X}(2-10~keV)\sim 10}$ \citep[see][]{Ho08}. 

\subsection{Alternative source of emission at mid-IR}

If faint LINERs are no longer dominated by the torus emission, what is the mechanism responsible for the mid-IR emission? This is a question for which we do not have a clear answer. Plausible contributors to the mid-IR emission are: (1) jet emission in the form of synchrotron radiation; (2) host galaxy contributors; and (3) advection-dominated accretion flows (ADAFs). Note that we have excluded the dust shell models because we do not see emission silicate features in the average mid-IR spectrum of faint LINERs, although it might be relevant for some of them \citep[e.g. NGC\,3998,][]{Mason13}. Fig. \ref{fig:JetElipAdaf} compares the shape of the mid-IR spectrum of faint LINERs with several mechanisms discussed along this section. Note that all of them are scaled to the emission at 15$\rm{\mu m}$. 

\subsubsection{Host galaxy contribution}

We discard star-formation as the dominant emission since the SB-dominated LINERs were not considered in the analysis, although it can still be present in a fraction as suggested by the PAH emission. 

Alternatively, it could be linked to the diffuse emission coming from the galaxy. LINERs are generally found in elliptical galaxies \citep[see][]{Carrillo99}. Indeed in very few cases faint LINERs are hosted in late type galaxies (see Section \ref{sec:Subclasses}). To compare the mid-IR spectra of LINERs with elliptical galaxies we have chosen the \emph{Spitzer}/IRS spectrum of the prototypical elliptical galaxy NGC\,1407\footnote{We have chosen NGC\,1407 because it is one of the few elliptical galaxies with full coverage of the mid-IR spectrum.}.   Although the general slope of faint LINERs resembles that of an elliptical galaxy, it lacks of PAH features and emission lines seen in faint LINERs.

\citet{Singh13} found that a large fraction of the optical spectrum of many LINERs could come from post- asymptotic giant branch (post-AGB) stars. In order to deeply investigate this, Fig.~\ref{fig:JetElipAdaf} (bottom) compares the spectrum of faint LINERs and SBs with different stages of the AGB stars.

Our average LINER faint spectrum shows PAH features (7.7, 8.6, 11.3 and 17 $\rm{\mu m}$) together with emission lines (e.g. [Ne II], [Ne III], [S III], and [O IV]). This is very different for AGB stars (see the spectrum of an AGB star SMP\,SMC\,18 in Fig.~\ref{fig:JetElipAdaf}, bottom) which show featureless continuum-dominated spectra \citep{Yang04,Sloan06}. Post-AGB stars (also called photo planetary nebulae, proto-PNe) show some nebular emission lines and/or PAH features (see the spectrum of post-AGB star IRAS\,01005+7910 in Fig.~\ref{fig:JetElipAdaf}, bottom). However, they show a peak on the continuum emission above 25$\rm{\mu m}$ and they lack of the [O IV] line at 26$\rm{\mu m}$ seen in faint LINERs \citep{Cerrigone09}.

The average spectrum of faint LINERs is more similar to evolved carbon-rich PNe, where the continuum is less prominent but PAH features and highly ionised nebular lines (e.g. [O IV]) are still present \citep{Stanghellini07}. Fig. \ref{fig:JetElipAdaf} (bottom) shows the comparison between the average spectrum for faint LINERs and the prototype carbon-rich PNe SMP\,LMC\,79. 
Although less prominent, the continuum of these carbon-rich PNe is still present. \citet{Stanghellini12} showed that the dust continuum of carbon-rich PNe is well fitted with a black-body model with a temperatures ranging $\rm{T\sim}$100-160 K. Moreover, at least 10$\rm{^6}$ PNe are needed if they are the only responsible for the mid-IR luminosity of faint LINERs, since PNe typically show IR luminosities of $\rm{L_{IR}\sim 1\times 10^{34}erg~s^{-1}}$ \citep{Stanghellini12}. Alternatively, a combination of elliptical galaxies (top panel of Fig. \ref{fig:JetElipAdaf}) and carbon-rich PNe could also reproduce the mid-IR average spectrum of faint LINERs. Finally, we have checked the [OI]/H$\rm{\alpha}$ ratio for our sub-sample of faint LINERs, and in eight out of the 22 of them the ratio is too high (i.e. [OI]/H$\rm{\alpha>0.25}$) to be explained by post-AGB stars \citep{Cid-Fernandes04}. Thus, none of these stages of AGB stars can fully accomplish for the observed features of faint LINERs, although a combination of several of them can explain the mid-IR spectrum for some of these faint LINERs.

Interestingly, a combination of carbon-rich post-AGB stars and PNe could explain the SB-dominated spectra. Moreover, the average spectrum of SBs SB-dominated S2s and SB-dominated LINERs show very similar spectra (see Fig. \ref{fig:SampleAGN}). Thus, in all these cases a combination of carbon-rich post-AGB and evolved PNe might be relevant at mid-IR. This is in agreement with the findings by \citet{Singh13}, where post-AGB stars dominate the optical spectrum when the AGN is faint.

\subsubsection{Advection-dominated accretion flows}

ADAF mechanisms have been largely claimed as the main responsible for the SED of LINERs and LLAGN in general \citep[e.g.][and references therein]{Nemmen14}. In these models the accretion disc is truncated at the inner parts and replaced by a hot, geometrically thick, optically thin accretion flow \citep{Narayan98}. This flow is radiatively inefficient accreting material to the inner parts. ADAF models are quite complex in showing many observational parameters. In order to compare our average mid-IR spectra with these models we have chosen the ADAF best-fit to NGC\,1097 reported by \citet{Nemmen14}. ADAF model (see Fig. \ref{fig:JetElipAdaf}) fails to reproduce the slope of the spectrum and also lacks of the PAH features and emission lines seen in faint LINERs.

\subsubsection{Jet emission}

LLAGN are generally radio loud according to their SED \citep{Ho08}. Several authors have pointed out that the full SED could be explained as emission from the jet \citep[e.g. NGC\,1052,][]{Fernandez-Ontiveros12}. \citet{Mason13} showed that jet emission is relevant for radio-loud LLAGN. To test this hypothesis, here we adopt a simplistic jet model based on internal shocks adapted for AGN \citep[see][and references therein]{Nemmen06,Nemmen14}. The parameters of the model are those used in \citet{Nemmen14}, assuming a power-law index distribution with index $\rm{p=2}$. Note that we have assumed that the optically thin part of the jet is the one dominating the mid-IR emission. Jet emission (similarly to the ADAF models) shows an spectral index which is opposite to that observed in faint LINERs (see top panel of Fig. \ref{fig:JetElipAdaf}). However,  a combination of jet and other mechanisms could also explain the average spectrum of faint LINERs.  


In summary, the shape of the mid-IR continuum of faint LINERs resembles that of elliptical galaxies, with a plausible contribution of carbon-rich PNe. However, it cannot be entirely described by any of the emission mechanism explained above. An AGN component might also be present at some level to account for lines like [O IV], which are sometimes suggested as indicative of AGN activity \citep[see][]{Dudik09}.  Note, however, that the emission mechanism producing the [O IV] line has been proposed to be unrelated to the AGN \citep{Pereira-Santaella10}. Indeed it is present in the spectrum of carbon-rich PNe (see the bottom panel of Fig. \ref{fig:JetElipAdaf}).

\section{Conclusions}\label{sec:conclusions}

We have analysed a sample of mid-IR spectra of 40 LINERs observed with \emph{Spitzer}/IRS. We compared the LINER sample with PG\,QSOs, Seyferts (S1 and S2), and Starburst (SB). The main results of this paper are: 

\begin{itemize}
\item We have developed a methodology to exclude SB-dominated mid-IR spectra based on the silicate optical depth $\rm{\tau_{9.7\mu m}}$, the strength of the $\rm{6.2\mu m}$ PAH feature, and the steepness of the mid-IR spectra. We have found that 25 out of the 40 LINERs do not show signatures of being SB-dominated. The fraction is similar to that obtained for Type-2 Seyferts.  
\item LINERs fall into the X-ray to mid-IR correlation for AGN only if Compton-thick candidates proposed at X-rays show an X-ray luminosity around 70 times higher than that computed at X-rays. This is expected if they are confirmed as Compton-thick AGN (see Section \ref{sec:XrayMIR}). Furthermore, four LINERs not previously classified as Compton-thick candidates are also consistent with being Compton-thick according to the X-ray to mid-IR relation. 
\item Bright LINERs (i.e. $\rm{L_{X}(2-10~keV)>10^{41} erg/s}$) tend to have an average mid-IR spectrum similar to that of Type-1 Seyferts. However, faint LINERs (i.e. $\rm{L_{X}(2-10~keV)<10^{41} erg/s}$) show a flatter average mid-IR spectrum, showing prominent emission lines. We suggest that this favours the disappearance of the dusty torus in LINERs with bolometric luminosities below $\rm{L_{bol}\simeq10^{42} erg/s}$, as predicted theoretically. Therefore, LINERs with bolometric luminosities below $\rm{L_{bol}\simeq10^{42} erg/s}$ might not longer be dominated by the torus in the mid-IR. Their mid-IR continuum emission resembles that of an elliptical galaxy although an AGN or a jet component together with some star-forming contribution (post-AGB stars and PNe) might also be present at some level.
\end{itemize}

A final caveat is that a large fraction of LINERs seem to be Compton-thick at X-rays (see previous subsection) although the torus emission have disappeared for them. How is that possible? A natural explanation is that the Compton-thick material seen at X-rays is not produced in the dusty torus. Dusty material absorbs continuum radiation both in the UV/optical and X-rays, and therefore the dusty torus might also be responsible for part of the X-ray obscuration. But dust-free gas attenuates just the X-ray continuum, so clouds inside the dust sublimation radius will provide additional obscuration only in this band. Conclusive evidence for such absorption comes from the short timescales for transit of X-ray absorbing clouds across the LOS, which establish the existence of obscuring clouds inside the dust sublimation radius \citep[e.g. NGC\,1365,][]{Risaliti09}. A natural explanation is that the Compton-thick clouds seen at X-rays in faint LINERs are produced in dust-free regions within the dust sublimation radius. Could LINERs be at a stage where the torus structure is already gone while the BLR is still present? This naturally explains why the optical spectrum of LINERs needs a population of post-main sequence stars, blocking the source of ionization that produces NLR lines \citep{Singh13}. Alternatively, this Compton-thick gas might not be related to the AGN, but rather to the host-galaxy as a result of galaxy interactions. In favour of that, many of the well known Compton-thick AGN are hosted in galaxy pairs or clusters \citep[e.g. NGC\,3690,][]{Goulding12,Gonzalez-Martin13} being even binary nuclei \citep[e.g. NGC\,6240,][]{Komossa03}. A closer look at the environment of Compton-thick AGN needs to be undertaken in order to get more light on this possibility. 

\begin{acknowledgements}
We thank to the referee for his/her useful comments and suggestions. Authors acknowledge Dr. Rodrigo Nemmen for his help on the SED models and to Dr. A. Garc\'ia-Hern\'andez for his advice on the mid-IR emission of AGB stars and PNe. Partially based on observations made with the Gran Telescopio Canarias (GTC), installed at the Spanish Observatorio del Roque de los Muchachos of the Instituto de Astrof\'isica de Canarias, in the island of La Palma.This research has been supported by the Spanish Ministry of Economy and Competitiveness (MINECO) under the grant (project refs. AYA2013-42227-P, AYA 2012-39168-C03-01, and AYA 2010-15169) and by La Junta de Andaluc\'ia (TIC 114). AAH acknowledges support from grant AYA2012-31447. DD acknowledges support from grant 107313 from PAPIIT, UNAM.  CRA is supported by a Marie Curie Intra European Fellowship within the 7th European Community Framework Programme (PIEF-GA-2012-327934).
\end{acknowledgements}


\onecolumn
\footnotesize{ 
\begin{landscape}
\begin{longtable}{r c c c l c c c c c c}
\hline \hline
 Object & Class & Thickness & $\rm{log(L_{X}})$ & $\rm{log(\nu L_{\nu} (12\mu m))}$  & $\rm{log(\nu L_{\nu}(20\mu m)/\nu L_{\nu}(30\mu m))}$ &  $\rm{\tau_{9.7}}$ & EW(PAH 6.2$\rm{\mu m}$) & EW(PAH 11.3$\rm{\mu m}$) & PAH-vs-$\rm{\tau_{9.7}}$ & Steepness \\
\hline
\endfirsthead 
\hline \hline
 Object & Class & Thickness & $\rm{log(L_{X}})$ & $\rm{log(\nu L_{\nu} (12\mu m))}$  & $\rm{log(\nu L_{\nu}(20\mu m)/\nu L_{\nu}(30\mu m))}$ & $\rm{\tau_{9.7}}$ & EW(PAH 6.2$\rm{\mu m}$) & EW(PAH 11.3$\rm{\mu m}$) & PAH-$\rm{\tau_{9.7}}$ & Steepness \\
\hline
\endhead
          NGC315 &   LINER1 &   NCT &   41.8 &   42.5 &   -0.070 &    0.070 $\rm{\pm}$    0.011 &    0.026 $\rm{\pm}$    0.010 &    0.073 $\rm{\pm}$    0.004 &        Weak&              -- \\ 
       IIIZw035 &   LINER2 &   CTc &   40.0 &   43.1 (43.3) &   -0.803 &    1.046 $\rm{\pm}$    0.061 &    0.339 $\rm{\pm}$    0.077 &    0.496 $\rm{\pm}$    0.042 &      Strong&     Steep \\ 
        NGC1052 &   LINER1 &   NCT &   41.2 &   42.3 (42.2) &    0.074 &    0.179 $\rm{\pm}$    0.000 &    0.022 $\rm{\pm}$    0.003 &    0.022 $\rm{\pm}$    0.001 &        Weak&              -- \\ 
        NGC1097 &   LINER1 &   NCT &   40.8 &   43.4 (41.3) &   -0.235 &   -0.158 $\rm{\pm}$    0.000 &    0.405 $\rm{\pm}$    0.001 &    0.624 $\rm{\pm}$    0.001 &      Strong&              -- \\ 
        NGC1291 &   LINER2 &   NCT &   39.0 &   41.6 &    0.007 &   -0.002 $\rm{\pm}$    0.000 &    0.033 $\rm{\pm}$    0.003 &    0.226 $\rm{\pm}$    0.005 &        Weak&              -- \\ 
        NGC2639 &   LINER1 &   CTc &   40.3 &   41.9 &   -0.220 &    0.628 $\rm{\pm}$    0.000 &    0.057 $\rm{\pm}$    0.018 &    0.303 $\rm{\pm}$    0.011 &        Weak&              -- \\ 
        NGC2655 &   LINER2 &   NCT &   41.2 &   42.0 (41.8*) &   -0.128 &    0.192 $\rm{\pm}$    0.019 &   -0.011 $\rm{\pm}$    0.009 &    0.190 $\rm{\pm}$    0.005 &        Weak&              -- \\ 
        NGC2685 &   LINER2 &   CTc &   39.0 &   40.6 (39.9*) &   -0.009 &    0.017 $\rm{\pm}$    0.000 &    0.026 $\rm{\pm}$    0.014 &    0.078 $\rm{\pm}$    0.010 &        Weak&              -- \\ 
       UGC04881 &   LINER2 &   CTc &  -38.4 &   43.7 &   -0.442 &    0.594 $\rm{\pm}$    0.013 &    0.696 $\rm{\pm}$    0.013 &    0.793 $\rm{\pm}$    0.006 &      Strong&     Steep \\ 
          3C218 &   LINER2 &   NCT &   42.1 &   43.0 &   -0.211 &    1.064 $\rm{\pm}$    0.000 &   -0.038 $\rm{\pm}$    0.043 &    0.572 $\rm{\pm}$    0.057 &        Weak&              -- \\ 
        NGC2841 &   LINER2 &   NCT &   39.2 &   41.9 &    0.044 &   -0.154 $\rm{\pm}$    0.000 &    0.037 $\rm{\pm}$    0.007 &    0.248 $\rm{\pm}$    0.008 &        Weak&              -- \\ 
        NGC3079 &   LINER2 &    CT &   42.1 &   42.5 &   -0.609 &    1.400 $\rm{\pm}$    0.021 &    0.434 $\rm{\pm}$    0.013 &    0.781 $\rm{\pm}$    0.038 &   Deep  &     Steep \\ 
        NGC3185 &   LINER2 &   CTc &   39.4 &   41.9 ($\rm{<41.7}$) &   -0.237 &    0.198 $\rm{\pm}$    0.026 &    0.525 $\rm{\pm}$    0.049 &    0.563 $\rm{\pm}$    0.011 &      Strong&              -- \\ 
        NGC3190 &   LINER2 &   NCT &   39.5 &   42.3 &   -0.064 &    0.107 $\rm{\pm}$    0.016 &    0.105 $\rm{\pm}$    0.007 &    0.708 $\rm{\pm}$    0.008 &        Weak&              -- \\ 
        NGC3628 &   LINER2 &   NCT &   39.9 &   41.9 ($\rm{<40.6}$) &   -0.616 &    2.110 $\rm{\pm}$    0.000 &    0.428 $\rm{\pm}$    0.004 &    0.783 $\rm{\pm}$    0.007 &   Deep  &     Steep \\ 
        NGC4125 &   LINER2 &   CTc &   38.7 &   42.1 &    0.119 &   -0.213 $\rm{\pm}$    0.000 &   -0.011 $\rm{\pm}$    0.008 &    0.093 $\rm{\pm}$    0.011 &        Weak&              -- \\ 
        NGC4261 &   LINER2 &   NCT &   41.1 &   41.8 (41.6) &    0.030 &    0.081 $\rm{\pm}$    0.013 &   -0.069 $\rm{\pm}$    0.004 &    0.062 $\rm{\pm}$    0.002 &        Weak&              -- \\ 
        NGC4321 &   LINER2 &   NCT &   40.5 &   41.1 &    0.019 &    0.115 $\rm{\pm}$    0.016 &    0.042 $\rm{\pm}$    0.003 &    0.262 $\rm{\pm}$    0.004 &        Weak&              -- \\ 
        NGC4374 &   LINER2 &   CTc &   39.5 &   41.4 ($\rm{<40.8}$) &   -0.004 &   -0.100 $\rm{\pm}$    0.000 &    0.024 $\rm{\pm}$    0.006 &    0.154 $\rm{\pm}$    0.008 &        Weak&              -- \\ 
        NGC4438 &   LINER1 &   CTc &  $\rm{<}$39.0 &  41.6 (40.8) &   -0.304 &    0.123 $\rm{\pm}$    0.008 &    0.206 $\rm{\pm}$    0.008 &    0.493 $\rm{\pm}$    0.004 &        Weak&     Steep \\ 
        NGC4450 &   LINER1 &   NCT &   40.3 &   41.6 &    0.142 &   -0.670 $\rm{\pm}$    0.000 &    0.119 $\rm{\pm}$    0.016 &    0.453 $\rm{\pm}$    0.012 &        Weak&              -- \\ 
        NGC4486 &   LINER2 &   NCT &   40.8 &   41.7 (41.1) &    0.075 &   -0.141 $\rm{\pm}$    0.000 &    0.028 $\rm{\pm}$    0.025 &   -0.010 $\rm{\pm}$    0.022 &        Weak&              -- \\ 
        NGC4552 &   LINER2 &   NCT &   39.2 &   41.7 &    0.196 &   -0.264 $\rm{\pm}$    0.000 &    0.051 $\rm{\pm}$    0.006 &    0.014 $\rm{\pm}$    0.008 &        Weak&              -- \\ 
        NGC4579 &   LINER1 &   NCT &   41.1 &   42.5 (41.9) &    0.002 & -136.483 $\rm{\pm}$    0.000 &    0.130 $\rm{\pm}$    0.008 &    0.259 $\rm{\pm}$    0.004 &        Weak&              -- \\ 
        NGC4589 &   LINER2 &   CTc &   38.9 &   41.2 &    0.256 &    0.221 $\rm{\pm}$    0.000 &    0.156 $\rm{\pm}$    0.009 &    0.197 $\rm{\pm}$    0.009 &        Weak&              -- \\ 
        NGC4594 &   LINER2 &   NCT &   40.0 &   41.8 (40.1) &    0.046 &    0.200 $\rm{\pm}$    0.000 &    0.030 $\rm{\pm}$    0.002 &    0.132 $\rm{\pm}$    0.002 &        Weak&              -- \\ 
       NGC4676A &   LINER2 &   NCT &   39.9 &   43.6 &   -0.397 &    0.909 $\rm{\pm}$    0.012 &    0.334 $\rm{\pm}$    0.009 &    0.747 $\rm{\pm}$    0.004 &      Strong&     Steep \\ 
        NGC4698 &   LINER2 &   CTc &   38.7 &   41.3 ($\rm{<40.9}$) &   -0.199 &   -0.222 $\rm{\pm}$    0.000 &    0.016 $\rm{\pm}$    0.016 &    0.028 $\rm{\pm}$    0.025 &        Weak&              -- \\ 
        NGC4696 &   LINER2 &   NCT &   40.0 &   41.5 &    0.224 &   -0.140 $\rm{\pm}$    0.000 &    0.010 $\rm{\pm}$    0.011 &    0.021 $\rm{\pm}$    0.014 &        Weak&              -- \\ 
        NGC4736 &   LINER2 &   NCT &   38.6 &   42.0 (40.0) &   -0.300 &   -0.359 $\rm{\pm}$    0.000 &    0.153 $\rm{\pm}$    0.001 &    0.680 $\rm{\pm}$    0.001 &        Weak&     Steep \\ 
        NGC5005 &   LINER1 &   CTc &  $\rm{<}$39.9 &   42.4 (41.5) &   -0.437 &    0.368 $\rm{\pm}$    0.046 &    0.200 $\rm{\pm}$    0.009 &    0.702 $\rm{\pm}$    0.014 &        Weak&     Steep \\ 
       MRK266NE &   LINER2 &   CTc &   41.7 &   43.3 (43.2*) &   -0.325 &    0.385 $\rm{\pm}$    0.066 &    0.405 $\rm{\pm}$    0.083 &    0.336 $\rm{\pm}$    0.022 &      Strong&     Steep \\ 
        NGC5866 &   LINER2 &   CTc &   38.3 &   42.0 ($\rm{<}$40.7) &   -0.144 &    0.189 $\rm{\pm}$    0.021 &    0.057 $\rm{\pm}$    0.005 &    0.518 $\rm{\pm}$    0.006 &        Weak&              -- \\ 
        NGC6240 &   LINER2 &    CT &   44.3 &   44.3 (43.5) &   -0.436 &    1.184 $\rm{\pm}$    0.039 &    0.331 $\rm{\pm}$    0.017 &    0.525 $\rm{\pm}$    0.017 &      Strong&     Steep \\ 
        NGC6251 &   LINER2 &   CTc &   41.6 &   42.9 (42.6*) &    0.124 &   -0.392 $\rm{\pm}$    0.000 &   -0.021 $\rm{\pm}$    0.003 &    0.025 $\rm{\pm}$    0.002 &        Weak&              -- \\ 
        NGC7130 &   LINER2 &    CT &   43.1 &   43.8 (43.2) &   -0.260 &    0.772 $\rm{\pm}$    0.000 &    0.240 $\rm{\pm}$    0.019 &    0.297 $\rm{\pm}$    0.020 &      Strong&     Steep \\ 
        NGC7331 &   LINER2 &   CTc &   40.5 &   42.3 &   -0.143 &    0.007 $\rm{\pm}$    0.007 &    0.124 $\rm{\pm}$    0.005 &    0.483 $\rm{\pm}$    0.003 &        Weak&              -- \\ 
         IC1459 &   LINER2 &   NCT &   40.5 &   41.6 ($\rm{<41.6}$) &   -0.127 &    0.213 $\rm{\pm}$    0.000 &    0.055 $\rm{\pm}$    0.004 &    0.108 $\rm{\pm}$    0.003 &        Weak&              -- \\ 
  NPM1G-12.0625 &   LINER2 &   NCT &  $\rm{<}$41.5 &   43.2 &   -0.243 &    1.051 $\rm{\pm}$   -0.000 &    0.006 $\rm{\pm}$    0.094 &    0.009 $\rm{\pm}$    0.042 &        Weak&     Steep \\ 
        NGC7743 &   LINER2 &   CTc &  $\rm{<}$39.5 &   41.5 ($\rm{<}$41.1) &   -0.252 &   -0.442 $\rm{\pm}$    0.000 &    0.054 $\rm{\pm}$    0.021 &    0.463 $\rm{\pm}$    0.015 &        Weak&     Steep \\ \hline
       Fairall9 &       S1 &   NCT &   44.0 &   44.6 (44.5)  &    0.201 &   -0.162 $\rm{\pm}$    0.000 &    0.003 $\rm{\pm}$    0.005 &    0.013 $\rm{\pm}$    0.004 &        Weak&              -- \\ 
        NGC526A &       S1 &   NCT &   43.2 &   43.6 (43.7) &    0.283 &   -0.096 $\rm{\pm}$    0.000 &   -0.008 $\rm{\pm}$    0.009 &    0.009 $\rm{\pm}$    0.007 &        Weak&              -- \\ 
        MRK1066 &       S2 &    CT &   42.9 &   43.5 &   -0.232 &    0.395 $\rm{\pm}$    0.033 &    0.256 $\rm{\pm}$    0.008 &    0.422 $\rm{\pm}$    0.009 &      Strong&              -- \\ 
        NGC1386 &       S2 &    CT &   41.6 &   42.4 (42.2) &   -0.138 &    0.622 $\rm{\pm}$    0.036 &    0.007 $\rm{\pm}$    0.006 &    0.057 $\rm{\pm}$    0.007 &        Weak&              -- \\ 
        NGC2110 &       S2 &   NCT &   42.4 &   43.1 (43.0) &   -0.010 &   -0.210 $\rm{\pm}$    0.000 &    0.015 $\rm{\pm}$    0.005 &    0.036 $\rm{\pm}$    0.006 &        Weak&              -- \\ 
      E005-G004 &       S2 &    CT &   41.9 &   42.2 (41.6) &   -0.242 &    1.199 $\rm{\pm}$    0.031 &    0.104 $\rm{\pm}$    0.012 &    0.291 $\rm{\pm}$    0.010 &        Weak&     Steep \\ 
           MRK3 &       S2 &    CT &   44.4 &   44.0 (43.7) &    0.103 &    0.547 $\rm{\pm}$    0.000 &    0.007 $\rm{\pm}$    0.006 &    0.009 $\rm{\pm}$    0.005 &        Weak&              -- \\ 
        NGC2273 &       S2 &    CT &   42.2 &   43.0 &   -0.133 &    0.235 $\rm{\pm}$    0.037 &    0.154 $\rm{\pm}$    0.008 &    0.271 $\rm{\pm}$    0.012 &        Weak&              -- \\ 
 IRAS07145-2914 &       S2 &    CT &   42.5 &   42.8 &   -0.028 &    0.120 $\rm{\pm}$    0.029 &    0.075 $\rm{\pm}$    0.009 &    0.201 $\rm{\pm}$    0.019 &        Weak&              -- \\ 
    MCG-5-23-16 &       S2 &   NCT &   43.0 &   43.4 (43.4) &    0.088 &    0.084 $\rm{\pm}$    0.026 &    0.005 $\rm{\pm}$    0.012 &    0.008 $\rm{\pm}$    0.011 &        Weak&              -- \\ 
        NGC3081 &       S2 &   NCT &   42.5 &   43.0 (42.8) &    0.019 &    0.052 $\rm{\pm}$    0.034 &    0.009 $\rm{\pm}$    0.010 &    0.041 $\rm{\pm}$    0.005 &        Weak&              -- \\ 
        NGC3281 &       S2 &    CT &   43.2 &   43.5 (43.5) &   -0.128 &    1.094 $\rm{\pm}$    0.033 &    0.020 $\rm{\pm}$    0.013 &    0.015 $\rm{\pm}$    0.020 &        Weak&              -- \\ 
        NGC3393 &       S2 &    CT &   42.9 &   42.9 (42.8) &   -0.100 &    0.617 $\rm{\pm}$    0.000 &    0.027 $\rm{\pm}$    0.016 &    0.084 $\rm{\pm}$    0.008 &        Weak&              -- \\ 
        NGC3621 &       S2 &   NCT &   39.3 &   41.6 &   -0.039 &    0.004 $\rm{\pm}$    0.011 &    0.434 $\rm{\pm}$    0.010 &    0.793 $\rm{\pm}$    0.005 &      Strong&              -- \\ 
        NGC3783 &       S1 &   NCT &   42.8 &   43.5 &    0.065 &    0.175 $\rm{\pm}$    0.000 &    0.000 $\rm{\pm}$    0.014 &    0.006 $\rm{\pm}$    0.005 &        Weak&              -- \\ 
        NGC4388 &       S2 &   NCT &   41.7 &   43.2 (42.8) &   -0.125 &    0.660 $\rm{\pm}$    0.028 &    0.051 $\rm{\pm}$    0.005 &    0.116 $\rm{\pm}$    0.007 &        Weak&              -- \\ 
        NGC4507 &       S2 &   NCT &   43.1 &   43.7 (43.7) &    0.017 &    0.148 $\rm{\pm}$    0.000 &    0.008 $\rm{\pm}$    0.006 &    0.036 $\rm{\pm}$    0.004 &        Weak&              -- \\ 
        NGC4725 &       S2 &   NCT &   38.9 &   41.6 &    0.277 &   -3.452 $\rm{\pm}$    0.000 &   -0.006 $\rm{\pm}$    0.008 &    0.520 $\rm{\pm}$    0.014 &        Weak&              -- \\ 
        NGC4939 &       S2 &    CT &   42.6 &   42.5 &    0.019 &    0.235 $\rm{\pm}$    0.000 &    0.048 $\rm{\pm}$    0.025 &    0.066 $\rm{\pm}$    0.015 &        Weak&              -- \\ 
        NGC4941 &       S2 &   NCT &   40.5 &   41.6 (41.6) &   -0.048 &    0.021 $\rm{\pm}$    0.048 &    0.028 $\rm{\pm}$    0.018 &    0.043 $\rm{\pm}$    0.014 &        Weak&              -- \\ 
        NGC4945 &       S2 &    CT &   42.3 &   42.1 (40.0) &   -1.018 &    2.958 $\rm{\pm}$    0.025 &    0.406 $\rm{\pm}$    0.009 &    0.599 $\rm{\pm}$    0.021 &   Deep  &     Steep \\ 
        NGC5033 &       S1 &   NCT &   41.1 &   42.9 (41.3) &   -0.118 &    0.172 $\rm{\pm}$    0.006 &    0.313 $\rm{\pm}$    0.004 &    0.618 $\rm{\pm}$    0.002 &      Strong&              -- \\ 
        NGC5135 &       S2 &    CT &   43.1 &   43.7 (43.2) &   -0.281 &    0.513 $\rm{\pm}$    0.023 &    0.292 $\rm{\pm}$    0.009 &    0.454 $\rm{\pm}$    0.008 &      Strong&     Steep \\ 
        NGC5194 &       S2 &   CTc &   40.9 &   42.3 &   -0.188 &    0.196 $\rm{\pm}$    0.003 &    0.318 $\rm{\pm}$    0.003 &    0.658 $\rm{\pm}$    0.001 &      Strong&              -- \\ 
    MCG-6-30-15 &       S1 &   NCT &   42.8 &   42.7 (42.8) &    0.115 &    0.147 $\rm{\pm}$    0.000 &    0.006 $\rm{\pm}$    0.008 &    0.030 $\rm{\pm}$    0.009 &        Weak&              -- \\ 
        IC4329A &       S1 &   NCT &   42.4 &   44.2 (44.2) &    0.182 &   -0.033 $\rm{\pm}$    0.000 &    0.006 $\rm{\pm}$    0.004 &    0.013 $\rm{\pm}$    0.009 &        Weak&              -- \\ 
        NGC5347 &       S2 &    CT &   42.4 &   43.0 (43.0) &    0.054 &    0.113 $\rm{\pm}$    0.019 &    0.037 $\rm{\pm}$    0.010 &    0.043 $\rm{\pm}$    0.006 &        Weak&              -- \\ 
       Circinus &       S2 &    CT &   41.9 &   42.8 (42.6) &   -0.373 &    0.997 $\rm{\pm}$    0.012 &    0.051 $\rm{\pm}$    0.003 &    0.044 $\rm{\pm}$    0.003 &        Weak&     Steep \\ 
        NGC5506 &       S2 &   NCT &   43.0 &   43.4 (43.2) &   -0.038 &    0.603 $\rm{\pm}$    0.025 &    0.012 $\rm{\pm}$    0.006 &    0.045 $\rm{\pm}$    0.004 &        Weak&              -- \\ 
        NGC5548 &       S1 &   NCT &   42.8 &   43.6 (43.3) &    0.097 &   -0.064 $\rm{\pm}$    0.000 &    0.014 $\rm{\pm}$    0.008 &    0.046 $\rm{\pm}$    0.005 &        Weak&              -- \\ 
        NGC5643 &       S2 &    CT &   42.6 &   42.3 (42.3) &   -0.183 &    0.100 $\rm{\pm}$    0.028 &    0.185 $\rm{\pm}$    0.011 &    0.243 $\rm{\pm}$    0.007 &        Weak&              -- \\ 
        NGC5728 &       S2 &    CT &   43.0 &   42.7 (42.1) &   -0.393 &    0.784 $\rm{\pm}$    0.051 &    0.228 $\rm{\pm}$    0.017 &    0.595 $\rm{\pm}$    0.013 &        Weak&     Steep \\ 
    ESO103-G035 &       S2 &   NCT &   43.4 &   43.8 (43.7) &    0.032 &    0.590 $\rm{\pm}$    0.029 &    0.005 $\rm{\pm}$    0.006 &    0.003 $\rm{\pm}$    0.006 &        Weak&              -- \\ 
      E138-G001 &       S2 &    CT &   42.5 &   43.6 (41.6) &    0.133 &    0.118 $\rm{\pm}$    0.000 &    0.012 $\rm{\pm}$    0.005 &    0.038 $\rm{\pm}$    0.021 &        Weak&              -- \\ 
      H1846-786 &       S1 &   NCT &   44.6 &   44.2 &   -0.035 &    0.029 $\rm{\pm}$    0.060 &    0.048 $\rm{\pm}$    0.023 &    0.039 $\rm{\pm}$    0.012 &        Weak&              -- \\ 
 IRAS19254-7245 &       S2 &    CT &   44.5 &   44.8 &   -0.236 &    0.649 $\rm{\pm}$    0.023 &    0.074 $\rm{\pm}$    0.009 &    0.131 $\rm{\pm}$    0.012 &        Weak&              -- \\ 
         MRK509 &       S1 &   NCT &   42.9 &   44.3 (44.2) &    0.108 &   -0.029 $\rm{\pm}$    0.000 &    0.019 $\rm{\pm}$    0.006 &    0.057 $\rm{\pm}$    0.005 &        Weak&              -- \\ 
        NGC7172 &       S2 &    CT &   42.7 &   43.0 (42.9) &   -0.213 &    2.069 $\rm{\pm}$    0.035 &    0.039 $\rm{\pm}$    0.006 &    0.174 $\rm{\pm}$    0.013 &   Deep  &              -- \\ 
        NGC7213 &       S1 &   NCT &   42.2 &   42.8 (42.6) &    0.205 &   -0.433 $\rm{\pm}$    0.000 &    0.007 $\rm{\pm}$    0.007 &    0.043 $\rm{\pm}$    0.005 &        Weak&              -- \\ 
        NGC7314 &       S2 &   NCT &   42.3 &   42.1 (41.9) &   -0.122 &    0.546 $\rm{\pm}$    0.056 &    0.019 $\rm{\pm}$    0.015 &    0.067 $\rm{\pm}$    0.019 &        Weak&              -- \\ 
    MCG-2-58-22 &       S1 &   NCT &   44.3 &   44.1 &    0.093 &   -0.024 $\rm{\pm}$    0.000 &    0.000 $\rm{\pm}$    0.011 &    0.019 $\rm{\pm}$    0.009 &        Weak&              -- \\ 
        NGC7582 &       S2 &    CT &   42.6 &   42.6 (42.8) &   -0.301 &    0.296 $\rm{\pm}$    0.040 &    0.284 $\rm{\pm}$    0.037 &    0.315 $\rm{\pm}$    0.013 &      Strong&     Steep \\ \hline
     PG0026+129 &    PG\,QSO &   NCT &   44.5 &   44.5 (44.6) &    0.322 &   -0.151 $\rm{\pm}$    0.000 &   -0.014 $\rm{\pm}$    0.012 &    0.008 $\rm{\pm}$    0.015 &        Weak&              -- \\ 
     PG0050+124 &    PG\,QSO &   NCT &   43.9 &   45.0 &    0.020 &   -0.224 $\rm{\pm}$    0.000 &    0.006 $\rm{\pm}$    0.005 &    0.013 $\rm{\pm}$    0.004 &        Weak&              -- \\ 
     PG0052+251 &    PG\,QSO &   NCT &   44.6 &   44.9 (44.7) &    0.328 &   -0.233 $\rm{\pm}$    0.000 &   -0.000 $\rm{\pm}$    0.025 &    0.022 $\rm{\pm}$    0.016 &        Weak&              -- \\ 
     PG0804+761 &    PG\,QSO &   NCT &   44.5 &   45.0 &    0.212 &   -0.359 $\rm{\pm}$    0.000 &   -0.005 $\rm{\pm}$    0.006 &   -0.000 $\rm{\pm}$    0.005 &        Weak&              -- \\ 
     PG0838+770 &    PG\,QSO &   NCT &   43.5 &   44.6 &   -0.002 &   -0.098 $\rm{\pm}$    0.000 &    0.007 $\rm{\pm}$    0.007 &    0.020 $\rm{\pm}$    0.012 &        Weak&              -- \\ 
     PG0844+349 &    PG\,QSO &   NCT &   43.7 &   44.2 (44.0) &    0.237 &   -0.333 $\rm{\pm}$    0.000 &    0.004 $\rm{\pm}$    0.008 &    0.014 $\rm{\pm}$    0.011 &        Weak&              -- \\ 
     PG1001+054 &    PG\,QSO &   NCT &   43.0 &   44.7 &    0.131 &   -0.126 $\rm{\pm}$    0.000 &    0.032 $\rm{\pm}$    0.013 &    0.039 $\rm{\pm}$    0.022 &        Weak&              -- \\ 
     PG1116+215 &    PG\,QSO &   NCT &   44.5 &   45.2 &    0.201 &   -0.174 $\rm{\pm}$    0.000 &   -0.015 $\rm{\pm}$    0.028 &    0.019 $\rm{\pm}$    0.021 &        Weak&              -- \\ 
     PG1126-041 &    PG\,QSO &   NCT &   43.1 &   44.3 &    0.055 &    0.174 $\rm{\pm}$    0.000 &   -0.007 $\rm{\pm}$    0.008 &    0.038 $\rm{\pm}$    0.052 &        Weak&              -- \\ 
     PG1211+143 &    PG\,QSO &   NCT &   43.7 &   44.8 &    0.224 &   -0.216 $\rm{\pm}$    0.000 &   -0.000 $\rm{\pm}$    0.008 &    0.002 $\rm{\pm}$    0.006 &        Weak&              -- \\ 
     PG1229+204 &    PG\,QSO &   NCT &   43.5 &   44.2 &    0.158 &   -0.055 $\rm{\pm}$    0.000 &   -0.012 $\rm{\pm}$    0.008 &    0.004 $\rm{\pm}$    0.014 &        Weak&              -- \\ 
     PG1244+026 &    PG\,QSO &   NCT &   43.2 &   43.8 &    0.022 &   -0.026 $\rm{\pm}$    0.000 &    0.029 $\rm{\pm}$    0.011 &    0.027 $\rm{\pm}$    0.009 &        Weak&              -- \\ 
     PG1307+085 &    PG\,QSO &   NCT &   44.1 &   44.9 &    0.268 &   -0.003 $\rm{\pm}$    0.000 &   -0.052 $\rm{\pm}$    0.056 &    0.012 $\rm{\pm}$    0.037 &        Weak&              -- \\ 
     PG1309+355 &    PG\,QSO &   NCT &   43.8 &   45.3 &    0.235 &   -0.295 $\rm{\pm}$    0.000 &    0.042 $\rm{\pm}$    0.061 &    0.018 $\rm{\pm}$    0.025 &        Weak&              -- \\ 
     PG1351+640 &    PG\,QSO &   NCT &   43.1 &   45.0 &    0.063 &   -0.493 $\rm{\pm}$    0.000 &    0.026 $\rm{\pm}$    0.010 &    0.030 $\rm{\pm}$    0.006 &        Weak&              -- \\ 
     PG1402+261 &    PG\,QSO &   NCT &   44.1 &   45.1 &    0.002 &   -0.172 $\rm{\pm}$    0.000 &    0.009 $\rm{\pm}$    0.013 &    0.017 $\rm{\pm}$    0.014 &        Weak&              -- \\ 
     PG1411+442 &    PG\,QSO &   NCT &   43.4 &   44.7 &    0.197 &   -0.125 $\rm{\pm}$    0.000 &    0.007 $\rm{\pm}$    0.007 &    0.012 $\rm{\pm}$    0.008 &        Weak&              -- \\ 
     PG1426+015 &    PG\,QSO &   NCT &   44.1 &   44.7 &    0.132 &   -0.180 $\rm{\pm}$    0.000 &    0.004 $\rm{\pm}$    0.007 &    0.024 $\rm{\pm}$    0.009 &        Weak&              -- \\ 
     PG1435-067 &    PG\,QSO &   NCT &   44.0 &   44.5 &    0.223 &   -0.106 $\rm{\pm}$    0.000 &    0.012 $\rm{\pm}$    0.017 &    0.006 $\rm{\pm}$    0.014 &        Weak&              -- \\ 
     PG1440+356 &    PG\,QSO &   NCT &   43.8 &   44.6 &   -0.027 &   -0.076 $\rm{\pm}$    0.000 &    0.034 $\rm{\pm}$    0.006 &    0.099 $\rm{\pm}$    0.013 &        Weak&              -- \\ 
     PG1448+273 &    PG\,QSO &   NCT &   43.3 &   44.2 &    0.177 &   -0.102 $\rm{\pm}$    0.000 &    0.019 $\rm{\pm}$    0.010 &    0.046 $\rm{\pm}$    0.006 &        Weak&              -- \\ 
     PG1501+106 &    PG\,QSO &   NCT &   43.7 &   44.2 &    0.142 &   -0.060 $\rm{\pm}$    0.000 &   -0.001 $\rm{\pm}$    0.008 &    0.005 $\rm{\pm}$    0.008 &        Weak&              -- \\ 
     PG1613+658 &    PG\,QSO &   NCT &   44.2 &   45.1 &    0.007 &   -0.114 $\rm{\pm}$    0.000 &    0.020 $\rm{\pm}$    0.008 &    0.044 $\rm{\pm}$    0.009 &        Weak&              -- \\ 
     PG1626+554 &    PG\,QSO &   NCT &   44.2 &   44.3 &    0.411 &   -0.426 $\rm{\pm}$    0.000 &   -0.009 $\rm{\pm}$    0.008 &    0.010 $\rm{\pm}$    0.016 &        Weak&              -- \\ 
     PG2130+099 &    PG\,QSO &   NCT &   43.5 &   44.7 (44.6) &    0.107 &    0.059 $\rm{\pm}$    0.000 &    0.007 $\rm{\pm}$    0.007 &    0.009 $\rm{\pm}$    0.007 &        Weak&              -- \\ 
     PG2214+139 &    PG\,QSO &   NCT &   43.7 &   44.3 &    0.247 &   -0.260 $\rm{\pm}$    0.000 &    0.003 $\rm{\pm}$    0.004 &    0.009 $\rm{\pm}$    0.008 &        Weak&              -- \\ \hline
         NGC520 &       SB &   NCT &   40.0 &   42.7 &   -0.592 &    1.379 $\rm{\pm}$    0.030 &    0.371 $\rm{\pm}$    0.017 &    0.776 $\rm{\pm}$    0.014 &   Deep  &     Steep \\ 
        NGC0855 &       SB &   NCT &   37.9 &   40.5 &   -0.366 &    0.369 $\rm{\pm}$    0.053 &    0.475 $\rm{\pm}$    0.101 &    0.549 $\rm{\pm}$    0.020 &      Strong&     Steep \\ 
        NGC0925 &       SB &   NCT &   38.3 &   41.0 &   -0.159 &    0.757 $\rm{\pm}$    0.000 &    0.454 $\rm{\pm}$    0.040 &    0.616 $\rm{\pm}$    0.015 &      Strong&              -- \\ 
          IC342 &       SB &   NCT &   39.0 &   41.4 &   -0.237 &    0.890 $\rm{\pm}$    0.000 &    0.414 $\rm{\pm}$    0.009 &    0.478 $\rm{\pm}$    0.005 &      Strong&              -- \\ 
        NGC1482 &       SB &   NCT &   39.4 &   43.2 &   -0.329 &    0.432 $\rm{\pm}$    0.003 &    0.468 $\rm{\pm}$    0.002 &    0.764 $\rm{\pm}$    0.001 &      Strong&     Steep \\ 
        NGC1614 &       SB &   NCT &   41.3 &   44.1 ($\rm{<43.7}$) &   -0.181 &    0.325 $\rm{\pm}$    0.051 &    0.445 $\rm{\pm}$    0.018 &    0.374 $\rm{\pm}$    0.023 &      Strong&              -- \\ 
        NGC1808 &       SB &   NCT &   39.7 &   42.6 (42.1) &   -0.295 &    0.132 $\rm{\pm}$    0.037 &    0.497 $\rm{\pm}$    0.009 &    0.677 $\rm{\pm}$    0.008 &      Strong&     Steep \\ 
        NGC2146 &       SB &   NCT &   39.0 &   42.6 &   -0.414 &    0.762 $\rm{\pm}$    0.038 &    0.544 $\rm{\pm}$    0.014 &    0.758 $\rm{\pm}$    0.008 &      Strong&     Steep \\ 
        NGC2798 &       SB &   NCT &   39.6 &   43.2 &   -0.297 &    0.008 $\rm{\pm}$    0.003 &    0.469 $\rm{\pm}$    0.005 &    0.599 $\rm{\pm}$    0.002 &      Strong&     Steep \\ 
        NGC2903 &       SB &   NCT &   39.9 &   42.4 &   -0.300 &    0.016 $\rm{\pm}$    0.000 &    0.495 $\rm{\pm}$    0.018 &    0.663 $\rm{\pm}$    0.007 &      Strong&     Steep \\ 
        NGC2976 &       SB &   NCT &   36.6 &   40.3 &   -0.075 &    0.885 $\rm{\pm}$    0.000 &    0.548 $\rm{\pm}$    0.036 &    0.612 $\rm{\pm}$    0.008 &      Strong&              -- \\ 
        NGC3184 &       SB &   NCT &   38.0 &   40.2 &   -0.138 &    1.220 $\rm{\pm}$    0.000 &   -0.151 $\rm{\pm}$    0.017 &    0.688 $\rm{\pm}$    0.015 &        Weak&              -- \\ 
        NGC3198 &       SB &   NCT &   38.2 &   41.9 &   -0.329 &    1.623 $\rm{\pm}$    0.008 &    0.255 $\rm{\pm}$    0.017 &    0.654 $\rm{\pm}$    0.009 &   Deep &     Steep \\ 
        NGC3256 &       SB &   NCT &   40.8 &   43.7 (43.6) &   -0.321 &    0.233 $\rm{\pm}$    0.035 &    0.473 $\rm{\pm}$    0.008 &    0.479 $\rm{\pm}$    0.014 &      Strong&     Steep \\ 
        NGC3310 &       SB &   NCT &   40.0 &   42.5 &   -0.359 &    0.029 $\rm{\pm}$    0.032 &    0.470 $\rm{\pm}$    0.017 &    0.614 $\rm{\pm}$    0.007 &      Strong&     Steep \\ 
        NGC3367 &       SB &   NCT &   40.9 &   42.9 &   -0.009 &    0.226 $\rm{\pm}$    0.000 &    0.251 $\rm{\pm}$    0.039 &    0.138 $\rm{\pm}$    0.011 &      Strong&              -- \\ 
           M108 &       SB &   NCT &   39.3 &   42.0 &   -0.489 &    0.272 $\rm{\pm}$    0.059 &    0.404 $\rm{\pm}$    0.028 &    0.618 $\rm{\pm}$    0.015 &      Strong&     Steep \\ 
          MRK52 &       SB &   NCT &   38.0 &   42.6 &   -0.074 &    0.633 $\rm{\pm}$    0.000 &    0.460 $\rm{\pm}$    0.033 &    0.259 $\rm{\pm}$    0.009 &      Strong&              -- \\ 
        NGC7252 &       SB &   NCT &   40.6 &   43.2 &   -0.318 &    0.274 $\rm{\pm}$    0.037 &    0.496 $\rm{\pm}$    0.019 &    0.830 $\rm{\pm}$    0.012 &      Strong&     Steep \\ 
\hline \hline
\caption{Sample and results. We have divided the sample according to their optical classification by horizontal lines. Columns show the following: 
(Col. 1) target name;
(Col. 2) object type; 
(Col. 3) Compton-thickness. CT: Compton-thick; NCT: No Compton-thick (i.e. Compton-thin); CTc: Compton-thick candidate using indirect arguments (see text).
(Col. 4) logarithmic of the intrinsic X-ray luminosity in the 2-10 keV band;
(Col. 5) logarithmic of the mid-IR luminosity obtained with \emph{Spitzer} at 12$\rm{\mu m}$. In parenthesis, the mid-IR luminosity obtained with ground-based observations is shown, when available. Those mir-IR luminosities with asterisks are presented here for the first time, using proprietary CanariCam/GTC observations;
(Col. 6) mid-IR steepness expressed as the ratio between the mid-IR luminosity at 20 and 30 $\rm{\mu m}$, $\rm{log(\nu L_{\nu}(20\mu m)/\nu L_{\nu}(30\mu m))}$; 
(Col. 7) silicate attenuation strength at 9.7$\rm{\mu m}$;
(Cols. 8 and 9) equivalent width of the PAH features at 6.2 and 11.3$\rm{\mu m}$;
(Col. 10) classification according to the EW(PAH 6.2$\rm{\mu m}$) and the $\rm{\tau_{9.7}}$. Weak: EW(PAH 6.2$\rm{\mu m)<0.233\mu m}$. Strong: EW(PAH 6.2$\rm{\mu m)>0.233\mu m}$. Deep: $\rm{\tau_{9.7\mu m}>1.25}$.
(Col. 11) classification according to the steepness of the mid-IR spectrum (i.e. $\rm{log(\nu L_{\nu}(20\mu m)/\nu L_{\nu}(30\mu m))}$). Steep: $\rm{log(\nu L_{\nu}(20\mu m)/\nu L_{\nu}(30\mu m))>-0.24}$.
}
\label{tab:sample}
\end{longtable}
\end{landscape}
}

\twocolumn

\end{document}